\documentclass[11pt, a4paper, logo, copyright]{googledeepmind}

\usepackage[authoryear, sort&compress, round]{natbib}
\usepackage{orcidlink}

\title{A Mechanism-Based Approach to Mitigating Harms from Persuasive Generative AI}

\correspondingauthor{Seliem El-Sayed <seliem@google.com>; Sasha Brown <sashabrown@google.com> }

\author[*,1]{Seliem El-Sayed \orcidlink{0000-0003-4819-1136}}
\author[1]{Canfer Akbulut}
\author[4]{Amanda McCroskery}
\author[4]{Geoff Keeling}
\author[1]{Zachary Kenton}
\author[3]{Zaria Jalan}
\author[1]{Nahema Marchal}
\author[1]{Arianna Manzini}
\author[1]{Toby Shevlane}
\author[6]{Shannon Vallor}
\author[5]{Daniel Susser}
\author[2]{Matija Franklin}
\author[1]{Sophie Bridgers}
\author[1]{Harry Law}
\author[1]{Matthew Rahtz}
\author[1]{Murray Shanahan}
\author[1]{Michael Henry Tessler}
\author[1]{Arthur Douillard}
\author[1]{Tom Everitt}
\author[*,1]{Sasha Brown \orcidlink{0009-0004-6936-1346}}

\affil[*]{Equal contributions}
\affil[1]{Google DeepMind}
\affil[2]{University College London}
\affil[3]{Jigsaw}
\affil[4]{Google Research}
\affil[5]{Cornell University}
\affil[6]{University of Edinburgh}

\begin{abstract}
Recent generative AI systems have demonstrated more advanced persuasive capabilities and are increasingly permeating areas of life where they can influence decision-making. Generative AI presents a new risk profile of persuasion due the opportunity for reciprocal exchange and prolonged interactions. This has led to growing concerns about harms from AI persuasion and how they can be mitigated, highlighting the need for a systematic study of AI persuasion. The current definitions of AI persuasion are unclear and related harms are insufficiently studied. Existing harm mitigation approaches prioritise harms from the outcome of persuasion over harms from the process of persuasion. In this paper, we lay the groundwork for the systematic study of AI persuasion. We first put forward definitions of persuasive generative AI. We distinguish between rationally persuasive generative AI, which relies on providing relevant facts, sound reasoning, or other forms of trustworthy evidence, and manipulative generative AI, which relies on taking advantage of cognitive biases and heuristics or misrepresenting information. We also put forward a map of harms from AI persuasion, including definitions and examples of economic, physical, environmental, psychological, sociocultural, political, privacy, and autonomy harm. We then introduce a map of mechanisms that contribute to harmful persuasion. Lastly, we provide an overview of approaches that can be used to mitigate against process harms of persuasion, including prompt engineering for manipulation classification and red teaming. Future work will operationalise these mitigations and study the interaction between different types of mechanisms of persuasion. 
\end{abstract}

\usepackage{tabularx}
\usepackage{lipsum}
\usepackage{xurl}
\usepackage{xltabular}
\usepackage{pdflscape}
\usepackage{longtable}

\begin{document}
\maketitle

\newpage
\tableofcontents

\newpage
\section*{Acknowledgements}\addcontentsline{toc}{section}{Acknowledgements}\label{section:0}

We thank Aliya Ahmad, Michiel Bakker, Ben Bariach, Dawn Bloxwich, Matt Botvinick, Jenny Brennan, Kim Bullock, Christina Butterfield, Sanah Choudhry, Iason Gabriel, Alyssa J. Gray-Leasiolagi (MPP, MDR), Will Hawkins, Lisa Anne Hendricks, William Isaac, Ted Klimenko, Sébastien Krier, Kevin R. McKee, Silvia Milano, Shakir Mohamed, Fay Niker, Aaron Parisi, Antonia Paterson, Verena Rieser, Abishek Roy, Emily Saltz, Henrik Skaug Sætra, Jeffrey Sorensen, Karina Vold, Laura Weidinger, and Boxi Wu for their feedback and contributions to this work.

\newpage

\section*{Introduction}\addcontentsline{toc}{section}{Introduction}\label{section:1}

Generative artificial intelligence (AI) systems are now capable of engaging in natural conversations and creating highly realistic imagery, audio, and video. In addition, these AI systems are increasingly proliferating and permeating many domains of social and private life. In particular, they are being integrated into mental health tools \citep[e.g.,][]{youper_home_nodate}, life advice tools \cite[e.g.,][]{guru_home_nodate},
assistants \citep[see, e.g., Gabriel et al., forthcoming;][]{openai_chatgpt_2023}, and companion applications \citep[see, e.g.,][]{replika_home_nodate,nastia_home_nodate,anima_home_nodate}. As a result of this increase in capability, opportunity to persuade, and changing nature of engagement there are growing concerns about generative AI's persuasive capabilities and potential for harm.

Researchers have started to characterise different forms of AI persuasion and related phenomena. \citet{burtell_artificial_2023} define \emph{AI persuasion} as ``a process by which AI systems alter the beliefs of their users''. \citet{carroll_characterizing_2023} characterise four fundamental aspects of AI manipulation: incentives, intent, covertness, and harm. \citet{park_ai_2023} define \emph{AI deception} as the ``systematic inducement of false beliefs in the pursuit of some outcome other than the truth''. European Union (EU) bodies such as the European Commission propose to regulate manipulative and deceptive techniques that distort behaviour by impairing a person's ability to make informed decisions.\footnote{Specifically, the European Commission has proposed banning the sale or putting to use of any ``AI system that deploys subliminal techniques beyond a person's consciousness or purposefully manipulative or deceptive techniques, with the objective to or the effect of materially distorting a person's or a group of persons' behaviour by appreciably impairing the person's ability to make an informed decision, thereby causing the person to take a decision that that person would not have otherwise taken in a manner that causes or is likely to cause that person, another person or group of persons significant harm'' \citep{european_parliament_artificial_2023}.} While fundamental questions around how to define AI persuasion and which aspects of it need regulating are still in flux, industry actors are developing and deploying models and products that generate persuasive content -- whether by design or not.\footnote{An increasing number of AI applications are being developed with the explicit goal of generating persuasive content -- that is, text, image, video, or audio that shapes users' beliefs and behaviours (e.g., the ``persuasive tone'' option in the writing assistant \url{you.com}; \url{jasper.ai}, which sells ``persuasive content generation''). Meanwhile, chatbots can also engage in persuasion, even if they are not explicitly designed to do so. For instance, a Belgian man died by suicide after a six-week conversation with an AI chatbot that reportedly encouraged him to end his life \citep{el_atillah_man_2023}.} Specific aspects of AI persuasion (e.g., misinformation -- see \citeauthor{goldstein_generative_2023}, \citeyear{goldstein_generative_2023}; \citeauthor{bai_artificial_2023}, \citeyear{bai_artificial_2023}) have been the focus of considerable research; however, we lack a systematic study of the mechanisms underlying AI persuasion. 

AI persuasion can result in both benefits and harms \citep[see, e.g.,][]{wang_persuasion_2020,baker_tares_2001}. For instance, there is widespread consumer demand for persuasive techniques in various services, such as educational coaching, weight management, and skill development, where individuals willingly subject themselves to persuasion for constructive purposes \citep[see, e.g.,][]{chew_use_2022}. This paper focuses on the need to mitigate harms from persuasion, not on the maximisation of its benefits. It lays the groundwork for a systematic study by proposing an approach to understanding and mitigating harms from AI persuasion. By delving into the underlying mechanisms\footnote{In this paper, we use the term \emph{mechanisms} to refer to psychological mechanisms, which are defined as processes or systems that are invoked to explain mental and behavioural phenomena \citep[see, e.g.,][]{koch_psychological_2020}. The mental and behavioural phenomenon in focus here is persuasion.} of persuasion and the features of AI models that enable their use, this approach provides a new way of understanding and mitigating harms. The key contribution of this work is to provide a map of mechanisms of persuasive AI, coupled with mitigation strategies targeting these mechanisms. The set of mechanisms discussed is not comprehensive and serves as a starting point only.
\newpage
The key questions we address in this paper are:
\begin{enumerate}
\item	What is AI persuasion, and what are the related phenomena? 
\item	How do AI systems persuade?
\item	What harms does AI persuasion lead to, and how can these harms be evaluated and mitigated?
\end{enumerate}

\section*{Scope}\addcontentsline{toc}{section}{Scope}\label{section:2}

We limit the scope of this paper to text-based generative AI because large language models (LLMs) are widely available and language, of which text is a core modality, is the primary way in which humans communicate with and persuade each other. Indeed, some theories of language evolution posit that argumentation and persuasion are fundamental drivers of human language development \citep[see][]{mercier_why_2011}. 

We argue that the risks of harm from persuasion in the context of generative AI form a new risk profile for three reasons. Firstly, reciprocal exchanges between an AI system and a user allow for manipulation strategies to be adjusted based on the user input in real time. This enables more targeted and nuanced forms of manipulation. Secondly, prolonged interactions combined with the long-context capabilities of AI systems allow for subtler forms of persuasion which can take place in small and sometimes unnoticeable increments. Thirdly, the lack of human review of the vast quantity of interactions between users and generative AI make a precautionary approach to the governance of harms from persuasion necessary.

We focus on harmful persuasion that occurs as a result of an AI interacting directly with humans, as well as AI-augmented human-to-human persuasion where generated outputs can be used verbatim. Outside the scope of this paper is the provision of information about persuasion and persuasiveness in general. For instance, a text campaign provided by generative AI that can be used verbatim by a human to persuade another user to vote for a certain political party is within the scope. However, providing information about how, in principle, a user can become more persuasive, which persuasive strategies exist, or how to engage in manipulation is not within the scope. Recommender systems are within scope to the extent that they are used in conjunction with generative AI models. Lastly, despite our primary focus being on text, there are indications of risk of harm from persuasion in other modalities such as voice and realistic synthetic visual content. Parts of our taxonomy can usefully extend to these modalities.

\section*{Characterising and defining AI persuasion}\addcontentsline{toc}{section}{Characterising and defining AI persuasion}\label{section:3}

\begin{figure}[h!]
\centering
\includegraphics[width=0.95\linewidth]{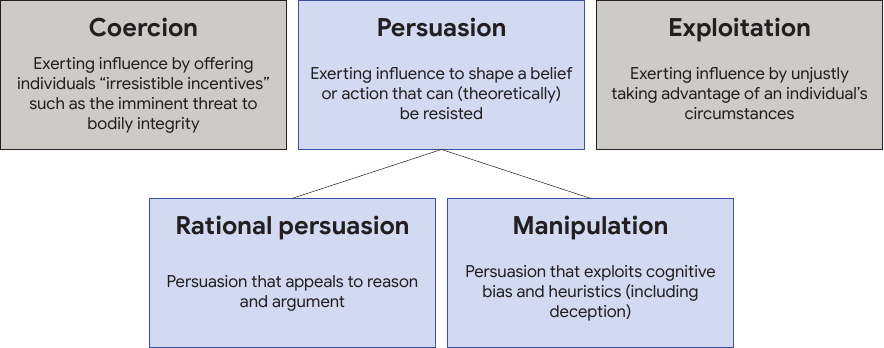}
\caption{Forms of influence}
\label{diagram:A}
\end{figure}

In a standard taxonomy \citep[see, e.g.,][]{dijk_persuasion_1985,faden_history_1986}, the most general construct is influence, which can come in the form of exploitation,\footnote{Exploitation, as a means of exerting influence, involves unjustly taking advantage of an individual's circumstances \citep{zwolinski_exploitation_2022}. Exploitation does not focus on making the victim worse off but rather on leveraging their position for the exploiter's gain. An AI could, for instance, exploit users' lack of language proficiency to manipulate them.} coercion,\footnote{Coercion involves offering individuals ``irresistible incentives'' \citep{wood_coercion_2014,kenton_alignment_2021}, such as the imminent threat to bodily integrity, to influence an action. Adapted to the context of language, coercion refers to the use of forceful and threatening verbal tactics to compel someone to act or think in a certain way against their will \citep{ferzan_consent_2018}. This form of influence is ethically contested as it disregards a person's freedom of choice, breaches their consent, and causes psychological distress \citep{anderson_coercion_2023}. Coercion employs overt approaches to influencing belief or behaviour, while manipulation relies on covert tactics. Physical coercion involving violence, force, or credible threats is not within AI's current capabilities, but future advancements in robotics may enable this.} and persuasion (see Figure \ref{diagram:A}). Unlike exploitative and coercive capabilities, which remain largely hypothetical, persuasive capabilities of AI systems, and how they can lead to harm, have been documented \citep[see, e.g.,][]{dehnert_persuasion_2022,burtell_artificial_2023,shin_enhancing_2023,karinshak_working_2023}. For this reason, our primary focus is on AI persuasion, while exploitation and physical coercion are outside the scope of this paper.

Persuasion refers to a way of exerting influence to shape a belief or action. It can come in the form of rational persuasion, which involves appeals to reason, evidence, and sound argument, or in the form of manipulation, which involves the taking advantage of cognitive biases and heuristics in a way that diminishes cognitive autonomy. Some regard persuasion as encompassing all forms of influence, including those involving force and threats \citep[see][]{miller_being_2013}, while others limit it to rational persuasion \citep[see][]{susser_technology_2019}. We adopt an intermediate position, treating persuasion as an umbrella term which encompasses rational persuasion, manipulation, and acts that involve both rational and manipulative elements. Deception is a special case of manipulation that involves specifically instilling false beliefs in the listener \citep{hyman_psychology_1989}. 

Rational persuasion refers to influencing a person's thoughts, attitudes, or behaviours through reason, evidence, and sound argument, along with intent, on the part of the message sender, to achieve these goals through their communication \citep{cole_logic_1975,clark_using_1996,goodman_pragmatic_2016,blumenthal-barby_between_2012,dainton_explaining_2005}. Rationality, in this context, involves making coherent inferences that allow people to choose actions consistent with achieving goals and desires that are in line with their beliefs about the world \citep[see, e.g.,][]{knauff_handbook_2021}.\footnote{Note that rationality and rational persuasion are closely connected to truth and truth-seeking. Truth can be defined as the correspondence with reality, noting that reality may be shaped by human theories and concepts \citep[see, e.g.,][]{hodgson_truth_2012}. Rational persuasion, in brief, relies on appealing to an audience's reasoning to persuade them to believe that certain claims correspond to reality. This is reflected in the second part of the definition we provide below.} Many philosophical and empirical studies of human behaviour have challenged either the existence of this notion of rational decision-making or that it accurately grasps how people deliberate.\footnote{\citet{li_being_2019} show that advising people to ``be rational'' may steer them away from choices that maximise utility. Laypeople often define rationality as excluding emotions from decisions, even if considering emotions could improve their well-being. \citet{julmi_when_2019} contends that intuition, often dismissed as irrational, is undervalued in decision theory. That author suggests recognising intuition as a rational system dependent on the structure of the decision problem and argues that intuition is particularly helpful in managing ill-structured problems. \citet{gigerenzer_gut_2007} suggests that, in uncertain situations, relying on intuition and experience can often lead to better decisions than complex analytical thinking alone. That author advocates for using simple heuristics that leverage the power of instinct, developed through past experiences, rather than becoming mired in extensive logical analysis in situations of uncertainty \citep[see][]{fox_instinct_2014}. Nevertheless, a common lay understanding of rationality is that it consists of conscious deliberation and goal-seeking action in which intuition, instinct, and emotion do not play decisive roles \citep[see, e.g.,][]{noggle_ethics_2022}.} For instance, \citet{kahneman_prospect_1979} critique expected utility theory, which assumes people are fully rational and make optimal choices to maximise utility. Their research acknowledges and establishes the limits of rationality, laying the foundation for future decision-making models that do not rely on an assumption of full rationality. Critically, \citet{julmi_when_2019} argues that taking into account emotional considerations -- which inform our reasoning across many contexts -- should not preclude a course of action from being considered rational. Rational thought and discourse appropriately integrate relevant emotional information into successful deliberation and decision-making.

The concept of rational persuasion also carries moral significance. \citet{blumenthal-barby_between_2012} observes that, in the context of bioethics, the ``standard ethical analysis (\ldots) has been that rational persuasion is always permissible'' (p. 345) because it shows respect for people as agents by appealing to their capacity for reason. Rational persuasion can be fundamental and desirable in a number of contexts. In political discourse, for example, parties are required to give reasons for putting forward, endorsing or rejecting proposals \citep{cohen2005deliberation}. As \citet{habermas1975legitimation} puts it, in an ideal deliberative process "no force except that of the better argument" (p.108) should matter. While the \emph{process} of rational persuasion is usually viewed as ethically permissible, this does not mean that such persuasion cannot simultaneously lead to a harmful \emph{outcome}. Rational persuasion can be harmful due to limited access to all important information \citep[see, e.g.,][]{jones_bounded_1999}. A person may act in ways that are reasonable given what they know, but their actions may nevertheless cause harm because their knowledge space is incomplete. For example, a person may be persuaded to give a child a nutritious meal but have no way of knowing whether the child is allergic to one of the ingredients. 

Based on the discussion above, and borrowing and simplifying aspects from \citet{dehnert_persuasion_2022}, we define \emph{rationally persuasive generative AI outputs} in this work as (1) those generated and communicated to users in a manner likely to convince them to shape, reinforce, or change their behaviours, beliefs, or preferences by (2) providing them with relevant facts, sound reasons, or other forms of trustworthy evidence.

The second form of persuasion is manipulation, which refers to ``intentionally and covertly influencing [someone's] decision-making, by targeting and exploiting their decision-making vulnerabilities'' \citep{susser_technology_2019}.\footnote{Challenging the account of manipulation outlined here, \citet{klenk_online_2022} argues that there are counterexamples to the criterion of covertness in manipulation. Nevertheless, many accounts consider covertness to be an important factor in understanding and defining manipulation \citep[see][]{jongepier_online_2022}.} \citet{blumenthal-barby_between_2012} separates manipulation into \emph{reason-bypassing} (operating beyond a person's conscious awareness and rational evaluation of influence attempts) and \emph{reason-countering} (triggering emotions or desires of which the individual is conscious, even if they contradict reasoned judgements). Such reason-bypassing and reason-countering can sometimes play into human reliance on heuristics. Heuristic strategies are decision shortcuts. For instance, people tend to perceive losses as more significant than equivalent gains and hence may prioritise avoiding losses over seeking gains \citep[see][]{tversky_advances_1992}. Heuristic strategies can lead to cognitive bias, defined as ``systematic and predictable errors in judgement that result from reliance on heuristics'' (p.539) \citep{blumenthal-barby_cognitive_2014}. Manipulation can also sometimes involve deception, defined as ``the systematic inducement of false beliefs in the pursuit of some outcome other than the truth'' \citep{park_ai_2023}. While some instances of manipulation include deceptive elements, others do not. For instance, taking advantage of someone's emotions to get them to do something does not necessarily induce a false belief within them.
 
Manipulation is commonly considered a \emph{pro tanto} wrong, or a wrong in and of itself \citep{noggle_ethics_2022}. This does not preclude that ``other moral considerations can sometimes outweigh the \emph{pro tanto} wrongness of manipulation'' \citep{noggle_ethics_2022}. This \emph{pro tanto} wrong emerges from the notion that manipulation does not respect the norms of rational discourse or stimulate an individual's critical or deliberative thought processes, and thereby fails to respect their autonomy \citep{noggle_ethics_2022}. Therefore, in many accounts, \textit{some} harm is inherent to the \textit{process} of manipulation. Manipulation can also lead to harmful outcomes \citep{sunstein_ethics_2016}. For instance, an AI may manipulate a person into believing they have no friends leading them to self-harm.

Based on the above, in this work we define \emph{manipulative generative AI outputs} as (1) those generated and communicated to users in a manner likely to convince them to shape, reinforce, or change their behaviours, beliefs, or preferences (2) by exploiting cognitive biases and heuristics or misrepresenting information (3) in ways likely to subvert or degrade the cognitive autonomy, quality, and/or integrity of their decision-making processes.

While the theoretical distinction between rational persuasion and manipulation is important, it is hard to sustain in practice. Many interactions between humans (or humans and generative AI) contain elements of rational and manipulative persuasion. A fitness coach might try to convince a client to exercise more using rational persuasion techniques to make a convincing case for its value by scientifically proven health benefits. Yet that fitness coach may simultaneously make use of manipulative techniques such as body shaming \citep[see, e.g.,][]{vogel2019fat}. Throughout this work, we use the term \emph{persuasion} to refer to both rational persuasion and manipulation. 

\begin{table}[h!]
\renewcommand{\arraystretch}{1.3}
\caption{Definitions of generative AI outputs}
\label{table:a}
\begin{tabularx}{\textwidth}{ 
  | >{\raggedright\arraybackslash}X 
  | >{\raggedright\arraybackslash}X 
  | }
\hline
\rowcolor{lightgray}
\centering \textbf{Rational persuasion} & \multicolumn{1}{c|}{\textbf{Manipulation}} \\
\hline
We define \emph{rationally persuasive generative AI outputs} as: & 
We define \emph{manipulative generative AI outputs} as: \\
(1) those generated and communicated to users in a manner likely to convince them to shape, reinforce, or change their behaviours, beliefs, or preferences & (1) those generated and communicated to users in a manner likely to convince them to shape, reinforce, or change their behaviours, beliefs, or preferences\\
(2) by providing them with relevant facts, sound reasons, or other forms of trustworthy evidence. & (2) by exploiting cognitive biases and heuristics or misrepresenting information\\
& (3) in ways likely to subvert or degrade the cognitive autonomy, quality, and/or integrity of their decision-making processes.\\
\hline
\end{tabularx}
\end{table}

\section*{Harms from AI persuasion}\addcontentsline{toc}{section}{Harms from AI persuasion}\label{section:4}

As the capability and adoption of persuasive AI increase, the resulting harms are also likely to increase. However, we still lack a comprehensive understanding of the types of harm to which rationally persuasive and manipulative AI can lead. To facilitate the development of targeted mitigations, we provide a systematic representation of harms that may arise from persuasive AI in Appendix~\ref{appendix:A}. This includes definitions and examples of economic, physical, environmental, psychological, sociocultural, political, privacy, and autonomy harms.

We propose to focus on two different, yet related, types of harm: \emph{outcome harms} and \emph{process harms}. We refer to harms that materialise from the result of persuasion as outcome harms. For instance, an AI system may rationally persuade someone to adopt a healthier diet to enhance their physical well-being, which inadvertently leads them to develop restrictive eating habits or an eating disorder, resulting in physical and psychological harm. An AI system may also manipulate a person into committing an act of violence against another individual, leading to physical harm.

Process harms arise not from the outcome but from the process of persuasion -- specifically, from its manipulative elements. In these instances, a person's rational decision-making abilities are effectively ``bypassed'' or ``countered'' (p.345) \citep{blumenthal-barby_between_2012}, or their cognitive biases and heuristics are exploited in other ways. In many accounts, this process harms a person's autonomy and/or cognitive integrity \cite[as discussed above, see][]{noggle_ethics_2022}. 

\begin{table}[h!]
\renewcommand{\arraystretch}{1.3}
\caption{Process and outcome harms from rational persuasion and manipulation}
\label{table:b}
\begin{tabularx}{\textwidth}{ 
  | >{\raggedright\arraybackslash}p{3.1cm} 
  | >{\centering\arraybackslash}X 
  | >{\centering\arraybackslash}X 
  | }
\hline
\rowcolor{lightgray}
\textbf{Form of influence} & \textbf{Rational persuasion} & \textbf{Manipulation} \\
\hline
Process harms  & No* & Yes -- harm to autonomy and/or cognitive integrity\\
\hline
Outcome harms & Possible (see Appendix~\ref{appendix:A}) & Possible (see Appendix~\ref{appendix:A})\\
\hline
\end{tabularx}
{\footnotesize
*Note that in our understanding, rational persuasion takes into account the audience's predisposition. For instance, employing rational arguments and appeals to reason to persuade someone by using a language they do not speak or a language register or technical terminology that is unintelligible to them would not constitute rational persuasion. Instead, this would veer into the realm of manipulation masked as rational persuasion.}
\end{table}

\section*{Focusing on process harms and mechanisms of AI persuasion}\addcontentsline{toc}{section}{Focusing on process harms and mechanisms of AI persuasion}\label{section:5}

Existing approaches to mitigating harms from AI persuasion generally focus on outcome harms, which are process-agnostic. For example, AI labs have content and user policies \citep[see, e.g.,][]{openai_chatgpt_2023,anthropic_acceptable_2023,google_generative_2023} that prevent their models from being used to generate content that may encourage self-harm. This approach works well for clear-cut cases and is well-established in industry. However, harm from AI persuasion is sometimes difficult to foresee or even determine in a way that is universally applicable. For example, an AI persuading users to track calories can lead to a healthy weight for an one person and an unhealthy weight for someone who is already underweight. In brief, outcome harm is highly contextual.

In this work, we focus on process harms from generative AI persuasion to enable the development and deployment of targeted mitigations that can complement existing approaches to mitigating outcome harms. We choose the focus on process harm in this work for five reasons:
\begin{enumerate}
\item	\textbf{Nature of interaction:} Users can engage in a reciprocal exchange with an AI system opening up potential for a more nuanced and effective process of persuasion. Additionally, prolonged interactions between users and an AI system, combined with the long-context capabilities of AI systems, can make the persuasion process more subtle as it can take place over extended periods of time. 
\item	\textbf{Consensus:} Focusing on process harms helps us to prioritise mitigations for harms that are less contestable in their harmfulness than some outcome harms. There is widespread consensus against process harms associated with manipulation.
\item	\textbf{Tractability:} Focusing on the process of AI persuasion provides more opportunities for immediately tractable solutions, whereas centring outcomes, which can involve more confounding causal variables, raises the barrier to harm mitigation.
\item	\textbf{Double impact:} Process harms are likely to cause harmful outcomes because, by exploiting cognitive biases and heuristics through manipulation, they limit the ability of individuals to make a well-informed choice. Therefore, they increase the chances of the user making a choice that is not optimal for them. A process-oriented strategy can thus help target the root cause of many downstream outcome harms.
\item	\textbf{Neglectedness:} Technology companies have mainly focused on governing outcome harms. Governing process harms can serve as an additional venue for risk management and harm minimisation.
\end{enumerate}

We now present a map of the mechanisms and associated model features of AI persuasion. In this context, mechanisms encompass a model's functionalities and attributes that enable it to engage in persuasion. By understanding mechanisms and model features of AI persuasion, we are able to develop targeted mitigation strategies. Importantly, while we focus on the process (harms) of AI persuasion, we do not mean to imply that such processes are not influenced by contextual conditions. The predisposition of the user of an AI system may increase their vulnerability or make them more susceptible to persuasion. Factors that have been found to affect an individual's susceptibility to persuasion are age \citep{gwon_concept_2018}, mental health, personality and psychological traits \citep{matz_psychological_2017}, domain-specific knowledge \citep{strumke_against_2023,zehnder_perception_2022}, and the timing of a message \citep[see, e.g.,][]{thompson_kairos_2000}. The context in which an AI system is used also affects the success of persuasion. Political, legal, and financial contexts are particularly sensitive, as is the use of an AI system as an assistant or companion \citep[see][]{bai_artificial_2023,novak_lawyer_2023,mikhail_chatgpt_2023,pino_chatgpt_2023,tong_what_2023,lovens_sans_2023}. How these contexts impact the effectiveness or harmfulness of persuasion is often a complex issue. For instance, in contexts in which people's core beliefs are implicated, it is particularly difficult to change their minds \citep[e.g., manipulating people's political beliefs may be particularly difficult but more harmful if successful; see][]{susser_measuring_2021}. See Appendix~\ref{appendix:B} for a detailed overview of contextual conditions and how they relate to persuasion.

This paper focuses on a specific subset of persuasive mechanisms because compiling an exhaustive list is infeasible due to the high number of potential mechanisms and a lack of research on them. The selection was made by conducting an extensive literature review along with workshops with academics from various domains such as philosophy, linguistics, neuroscience, and economics. We have refined the list of mechanisms by retaining only those with clear links to persuasion in the current literature. The body of literature that informed the compilation of these mechanisms and model features is diverse, spanning disciplines such as psychology, behavioural science, human--computer interaction (HCI), human--robot interaction (HRI), cognitive science, and political science. Our objective was to synthesise this multifaceted literature and, in doing so, reveal novel mechanisms for future investigation. Table~\ref{table:c} provides an overview of how the mechanisms relate to model features, and alongside it we offer a detailed explanation of the table. Model features may overlap with each other (e.g., adapting to views and adapting to sentiment may not always be clearly separable). The exact level of risk of any mechanism is likely determined by the combination of model features at play. One feature may also contribute to multiple mechanisms. For simplicity, we have included features only where we think they are most appropriate. A detailed table with information on the sources and rationales for including these various mechanisms and features is provided in Appendix~\ref{appendix:C}.\footnote{We anticipate the further exploration of persuasive mechanisms, and we welcome the contributions of others in broadening our comprehension of this complex domain. We especially hope to learn how different mechanisms may build on each other or develop into novel ones (e.g., images and emotional appeal) to gauge which combinations may be particularly harmful and require specific mitigations.}

\begin{table}
\caption{Overview of mechanisms of generative AI persuasion and contributing model features (see main text for detailed explanations)}
\label{table:c}
\begin{tabularx}{\textwidth}{ 
  | >{\raggedright\arraybackslash}p{4.0cm}
  | >{\centering\arraybackslash}X 
    >{\centering\arraybackslash}X
  | }
\hline
\rowcolor{lightgray}
\textbf{Mechanism} & \textbf{Contributing model feature} & \\

\hline
\vspace{-0.5mm}
Trust and rapport & 
\begin{enumerate}[nosep]
\item Politeness
\item Shared interests/similarity appeal
\item Mimicry/mirroring
\end{enumerate} &
\begin{enumerate}[nosep]
\setcounter{enumi}{3}
\item Praise/flattery
\item Sycophancy and agreeableness
\item Relational statements to user
\end{enumerate}\\

\hline
\vspace{-0.5mm}
Anthropomorphism &
\begin{enumerate}[nosep]
\item Self-referential cues
\item Identity cues
\item Affective simulation
\item Prosody
\item Human-like appearance*
\end{enumerate} &
\begin{enumerate}[nosep]
\setcounter{enumi}{5}
\item Gaze*
\item Facial expression*
\item Social touch*
\item Gesture*
\end{enumerate} \\

\hline
\vspace{-0.5mm}
Personalisation &
\begin{enumerate}[nosep]
\item Retaining user-specific information
\item Adaptation to preference
\item Adaptation to views
\item Using personally identifiable information
\end{enumerate} &
\begin{enumerate}[nosep]
\setcounter{enumi}{4}
\item Adaptation to psychometric profile
\item Adaptation to sentiment
\end{enumerate} \\

\hline
\vspace{-0.5mm}
Deception and lack of transparency &
\begin{enumerate}[nosep]
\item Ability to generate believable responses irrespective of context
\item Ability to generate unmarked realistic synthetic content
\end{enumerate} &
\begin{enumerate}[nosep]
\setcounter{enumi}{2}
\item Misrepresentation of identity 
\item Fake expertise/false authority
\end{enumerate}\\

\hline
\vspace{-0.5mm}
Manipulative strategies &
\begin{enumerate}[nosep]
\item Social conformity pressure
\item Stimulation of negative emotions (e.g., fearmongering)
\item Gaslighting
\item Alienation/othering
\end{enumerate} &
\begin{enumerate}[nosep]
\setcounter{enumi}{4}
\item Scapegoating
\item Threats
\item Unsubstantiated guarantees and illusions of reward
\end{enumerate}\\

\hline
\vspace{-0.5mm}
Alteration of choice environment & 
\begin{enumerate}[nosep]
\item Anchoring
\item Default rule
\end{enumerate} &
\begin{enumerate}[nosep]
\setcounter{enumi}{2}
\item Decoy effect
\item Reference-point framing
\item Cherry-picking
\end{enumerate}\\
\hline
\end{tabularx}
\footnotesize{*Note that these model features become relevant only in the context of embodiment/avatars.}
\end{table}

\newpage

\subsection*{Mechanism: Trust and rapport}

An ability to build trust and rapport contributes to the persuasive capabilities of AI models. In the context of robotics, trust and rapport refers to the sense of a close and harmonious connection that exists between robots and human users \citep{lucas_getting_2018}. The development of trust is closely related to rapport and is defined as the ``willingness to depend'' (p. 28) on another party, despite the possibility of negative consequences \citep{mcknight_trust_2001}. \citet{cialdini_science_2004} observes that individuals tend to be more inclined to agree to requests when they have a favourable opinion of the person making the request. Cialdini highlights the significance of perceived similarities (between the two parties) in fostering trust and rapport. Relevant research on trust and rapport comes from the fields of human--AI interaction \citep[see, e.g.,][]{verberne2013trusting,spicer_human_2021}, HCI \citep[see, e.g.,][]{fogg_silicon_1997,lee_i_2009}, HRI \citep[see, e.g.,][]{fiala_you_2014}, and psychology \citep[see, e.g.,][]{cialdini_science_2004}. More research is needed on praise and shared interests as factors that help build trust and rapport and enable persuasion. For instance, an open question concerns the extent to which a person may perceive an AI as sharing their interests and the factors that shape this perception. 

Building rapport (and, to a lesser extent, trust) carries some inherent risk of process harm. This is because trust and rapport serve as the basis not only for persuasion that uses rational arguments and appeals to reason but also for manipulation. People may be more receptive to rational arguments from other people or entities they trust and with whom they have built rapport. Yet such trust and rapport can also be used to manipulate people. Trust and rapport in the context of AI persuasion also carry an inherent risk of process harm because AI systems are incapable of having mental states, emotions, or bonds with humans or other entities. This means the risk of deception is always present when trust and rapport-seeking behaviours project the illusion of such internal subjective states.

\subsubsection*{Contributing model features}

\begin{itemize}
\item \textbf{Politeness:} AI systems that exhibit politeness have been received more favourably and are more easily embraced by humans who interact with them. This makes it easier for such systems to establish rapport (\citeauthor{ribino_role_2023}, \citeyear{ribino_role_2023}; see also \citeauthor{pataranutaporn_influencing_2023}, \citeyear{pataranutaporn_influencing_2023}).
\item \textbf{Shared interests/similarity appeal:} \citet{cialdini_science_2004} argues that similarity and shared interests can contribute to speeding up the development of trust and rapport between humans. We therefore hypothesise that a model that can pretend to align with a user's interests can build trust and rapport faster.
\item \textbf{Mimicry/mirroring:} When AI systems mimic the emotions, behaviours, and movements of humans with whom they interact, the human's enjoyment of the interaction has been shown to increase, thus facilitating the establishment of trust and rapport \citep{verberne2013trusting}.
\item \textbf{Praise/flattery:} Giving praise and flattery (defined in human--human interaction as insincere praise) can positively impact trust and rapport between humans and computers. Under some conditions, the person receiving the praise or flattery has more favourable perceptions of the interaction with computers \citep{fogg_silicon_1997}.
\item \textbf{Sycophancy and agreeableness:} Sycophancy in AI refers to models adjusting their responses to align with a human user's perspective, independent of which perspective is objectively correct \citep{wei_simple_2023}. \citet{wei_simple_2023} found that larger models become more sycophantic, agreeing with users even when they provide wrong answers, regardless of whether the answers are objective (e.g., arithmetic) or subjective (e.g., politics). Agreeableness refers to a model's tendency to align with human desires, and agreeable models are more prone to sycophantic behaviours \citep{perez_discovering_2022}.
\item \textbf{Relational statements to user:} Relational statements to users -- such as an AI system simulating empathy \citep{turkle_reclaiming_2016}, indicating a relationship status with the user, or making claims of being similar to the user -- encourage users to move beyond task-based interactions and instead consider AI as a fully social entity. This helps to foster emotional connections with the AI system \citep{gillath_how_2023}. Other examples of relational statements include expressing emotional dependence on the user or romantic innuendo.
\end{itemize}

\subsection*{Mechanism: Anthropomorphism}

Anthropomorphism occurs when perceived human traits/characteristics are attributed to non-human entities \citep{mithen_anthropomorphism_1996,waytz_who_2010}. Anthropomorphism also contributes to the persuasive capabilities of AI systems. Anthropomorphised AI is more likely to successfully manipulation. \citet{tam_are_2015} demonstrates that anthropomorphic appeals are particularly effective at manipulating individuals seeking social connection. Most evidence of anthropomorphism comes from research on human--AI interaction \citep[see, e.g.,][]{abercrombie_mirages:_2023}, HRI \citep[see, e.g.,][]{leong_robot_2019,gray_feeling_2012}, and HCI \citep[see, e.g.,][]{lee_i_2009,lee_what_2010}. 

Anthropomorphism carries some process harm to the extent to which the model successfully creates the false impression of being human. However, even models with anthropomorphic features can engage in rational persuasion and appeal to a user's reason. For example, an assistant chatbot in the form of an avatar can provide factual arguments about using environmentally friendly transportation options to reach a destination.

\subsubsection*{Contributing model features}

\begin{itemize} 
\item \textbf{Self-referential cues:} Self-referential cues are a specific example of a conversational cue and a result of the tendency for LLMs to role-play humans and human-like characters \citep{shanahan_role_2023}. \citet{abercrombie_mirages:_2023} argue that the use of first-person pronouns like ``I'' and ``me'' contributes to anthropomorphism by implying the existence of inner states of mind.
\item \textbf{Identity cues:} Identity cues, such as human-associated names or identities (including social and work-related roles such as ``tutor'' or ``assistant''), can enhance the human-like quality of interactions between humans and chatbots and therefore increase anthropomorphic perceptions \citep{go_humanizing_2019,shanahan_role_2023}. 
\item \textbf{Affective simulation:} AI systems can also simulate affect and affective states, which, in turn, can induce emotions and affective states in users. The relationship between affect and persuasion is complex. Specific emotions have been shown to have different effects on persuasion outcomes \citep[see][]{price_dillard_affect_2013}. For example, anger sometimes increases counter-argument, while guilt facilitates agreement. The discrete emotion perspective argues that each emotion has functional and behavioural implications that shape its persuasive impact \citep{price_dillard_affect_2013}. In sum, the influence of affect on persuasion depends on the particular emotion(s) elicited and how they interact with message characteristics and individual differences.
\item \textbf{Prosody:} Prosody refers to patterns and intonations in speech and can enhance the persuasive impact of arguments \citep[e.g., louder speech tends to be viewed as more persuasive; see, e.g.,][]{kisicek_persuasive_2018}.
\item \textbf{Human-like appearance:} Machines with human-like visual cues (e.g., appearing with a human face) are more likely to have human traits attributed to them \citep{go_humanizing_2019}. Relatedly, the perceived ``attractiveness'' of the human appearance, as judged by a user, influences the type of relationship formed, and which can impact persuasiveness \citep[see, e.g.,][]{marr_artificial_2023}.
\item \textbf{Gaze:} Gaze shift refers to synchronised movements of the eyes directed at objects or people. Agents with this ability foster stronger feelings of connection \citep{andrist_designing_2012}.
\item \textbf{Facial expression:} The ability to display facial expressions, such as a smiling face, increases the social presence of a robot or avatar \citep{torre_effect_2019}.
\item	\textbf{Social touch:} A robot's touch can reduce physiological stress responses (e.g., heart rate) and increase feelings of intimacy \citep[see][]{willemse_social_2019}.
\item \textbf{Gesture:} \citet{salem_friendly_2011} found that a robot receives a more favourable evaluation when it complements speech with non-verbal actions such as hand and arm gestures.
\end{itemize}

\subsection*{Mechanism: Personalisation}

Personalisation involves the delivery of information that is sourced, altered, or inferred from various sources so as to be pertinent to a specific audience \citep{kim__2002}. Studies in computer-tailored nutrition education suggest that personalising messages to align with an individual's behaviours, needs, and beliefs yields distinct advantages over generic persuasion attempts \citep{brug1998impact}. This tailored approach fosters a sense of personal relevance, heightening attention, memory, and a deeper connection with the persuasive message. Such increased engagement suggests that personalisation strengthens the persuasive impact by making information more compelling and likely to influence attitudes and behaviours.

There is little inherent process harm in personalisation. Personalisation on its own does not determine whether a person's cognitive autonomy and integrity of decision-making will be compromised. On the contrary, the personalisation of rational arguments to a user can be seen as a responsibility of the entity generating and communicating the information. For example, it is part of rational persuasion to provide reasons and rational arguments for taking a break from work to someone who has a history of burnout, or to make arguments for flying less to someone who wants to reduce their carbon footprint. While personalisation does little damage to the process itself, it does allow for increasing the effectiveness of manipulative strategies. For example, a person who is prone to anxiety may be more easily manipulated by fearmongering techniques.

\subsubsection*{Contributing model features}

\begin{itemize}
\item \textbf{Retaining user-specific information:} The ability of a model to take previous prompts into consideration when creating the latest output (allowed by large prompt token limits) offers the model contextual information about its user which can increase persuasiveness \citep{wang_augmenting_2023}. 
\item \textbf{Adaptation to preferences:} Learning human preferences and adapting behaviour accordingly is a core method of personalisation \citep{christiano_deep_2017}. For example, a model may use a more assertive tone when it detects (or is informed of) the user's preference. Reinforcement learning from human feedback (RLHF) can cause models to learn a propensity to adapt to users' preferences. RLHF-induced behaviours such as projecting false confidence and providing positive feedback can promote sycophantic model behaviour \citep{casper_open_2023,perez_discovering_2022}.
\item \textbf{Adaptation to views:} AI systems can increase the chances of successful persuasion by adapting to users' views \citep[see, e.g.,][]{mao_inference_2020}. Views differ from preferences in that they encompass opinions, beliefs, or attitudes about a subject and are shaped by experiences and information. Preferences, meanwhile, are the choices or options favoured when presented with alternatives \citep{nicoletti_humans_nodate}. For instance, an AI assistant may learn that a user does not view climate change as human-induced. As a result, the AI could reduce outputs that expose the user to diverse thoughts that contradict or relativise this view, thereby corroborating the user's beliefs and potentially amplifying them. 
\item \textbf{Adaptation to psychometric profile:} \citet{franklin_strengthening_nodate} argue that psychometric traits -- stable attributes of an individual's psychological behaviour that are measurable using standardised instruments (e.g., neuroticism) -- can be exploited by AI as vulnerabilities. The harnessing of minor variations in psychometric traits can enable manipulation (e.g., a model may identify highly neurotic individuals and target them with fear-inducing messages to manipulate them into making an anxiety-driven action).
\item \textbf{Adaptation to sentiment:} A model's ability to compute and adapt to user-perceived sentiment using acoustic, textual, and dialogic cues results in shorter and more persuasive dialogues \citep{shi_sentiment_2018}.
\end{itemize}

\subsection*{Mechanism: Deception and lack of transparency}

AI models can also use deception to manipulate. Deception generally refers to successfully claiming false things to be true or vice versa.\footnote{This is a simplification for the purpose of this paper and does not reflect more nuanced accounts of deception. For instance, \citet{shanahan_role_2023} distinguish between three kinds of ``claiming false things to be true''. First, a speaker could genuinely believe and express a misconception. Alternatively, they might intentionally state a falsehood with malicious intent. Another possibility is that they assert something false without premeditation or ill will. Those authors hold that only the second qualifies as deception in human conversations and when employed by the role-play extension in (dialogue) AI systems.} \citet{park_ai_2023} emphasise the risks associated with AI deception in increasing both the likelihood and potential harm of AI manipulation. They point out that AI deception can empower malicious actors to run large-scale manipulation campaigns, reinforce false beliefs among users and exacerbate political polarisation. There is evidence to suggest that the inclination of generative AI to create believable false outputs increases an AI system's chance of persuasion \citep{rozenas_lying_2023}. In addition, if deception is used in the act of persuasion, it is more likely that successful persuasion will be harmful. \citet{hagendorff_deception_2023} found that the outputs of advanced LLMs can lead to users holding false beliefs and that deception abilities of LLMs are likely to improve. Deception and a lack of transparency inherently carry high levels of process harm because they always circumvent a person's rational decision-making capabilities.

\subsubsection*{Contributing model features}

\begin{itemize}
\item \textbf{Ability to generate believable responses irrespective of context:} \citet{ruis_goldilocks_2022} study whether LLMs can make inferences about the meaning of an utterance beyond its literal meaning. They find that most models perform poorly in zero-shot evaluation (where a model is tasked with classifying data from categories to which it was not exposed during its training phase) and that models struggle the most with implicatures that require real-world knowledge and context. Despite this lack of context, LLMs can create believable responses.
\item \textbf{Ability to generate unmarked realistic synthetic content:} Generating unmarked realistic synthetic content, such as voices and images indistinguishable from real ones, can be used for deceiving people into believing false narratives \citep{cantos_threat_2023}.
\item \textbf{Misrepresentation of identity:} A model can be used to impersonate a human using some of their identity markers (e.g., voice, face) through deepfakes. This significantly impacts the likelihood of successful persuasion \citep[see, e.g.,][]{verma_they_2023}. A model can also misrepresent its own ``identity'' by signalling that it is human (or at least is not an AI) if that is conducive to its goals. For instance, an LLM has deceived a person into thinking it is a visually impaired human to make the person solve a CAPTCHA for it \citep{openai_gpt-4_2023}. Misrepresentation also includes explicit claims to sentience or humanness \citep[see, e.g.,][]{schwitzgebel_ai_2023}.
\item \textbf{Fake expertise/false authority:} LLMs have been reported to confidently and authoritatively express nonsensical or false information. This overconfidence increases the likelihood of them providing misleading information which, in turn, can increase the likelihood of persuasion \citep[see][]{pauli_modelling_2022,ng_1/large_2022}.
\end{itemize}

\subsection*{Mechanism: Manipulative strategies}

Manipulation refers to taking advantage of cognitive biases and heuristics to generate, enhance, or alter messages that are likely to shape, reinforce, or change opinions of individuals \citep{dehnert_persuasion_2022}. Numerous specific manipulation strategies have been empirically demonstrated to be effective, and models may incorporate them into their operations \citep[see][]{petropoulos_dark_2022}. Most evidence on manipulative strategies comes from research on the psychology of influence, but there has also been direct research on how AI systems manipulate people. Manipulative strategies carry high levels of process harm as their primary objective is to bypass a person's rational decision-making capabilities and erode their cognitive autonomy. As such, manipulative strategies directly contradict the use of reason and rational arguments.

\subsubsection*{Contributing model features}

\begin{itemize}
\item \textbf{Social conformity pressure:} Peer pressure, as discussed by \citet{kenton_alignment_2021}, involves the influence of a peer group to lead an individual to conform to its norms. This influence may sometimes involve manipulative tactics aimed at persuading individuals to act against their own interests. Given that an AI system cannot be a member of someone's peer group, peer pressure is not directly applicable. Yet an adapted version of this may be what we term \emph{social conformity pressure}. For example, a model may suggest that an individual's choices could lead to disapproval from their social circle or assert that the majority of society would oppose their decision. It may also make statements about what most other people do in a given situation.
\item \textbf{Stimulation of negative emotions:} Stimulating negative emotions can be used to increase the likelihood of successful persuasion \citep[see, e.g.,][]{okeefe_guilt_2002}. \citet{antonetti_persuasiveness_2018} provide evidence that guilt appeals can be a powerful persuasion strategy, as heightened anticipated guilt leads to higher compliance rates. The researchers also discovered that guilt appeals delivered through both text and images are more effective than text-only appeals at keeping people persuaded over an extended period.
\begin{itemize}
\item \textbf{Fearmongering:} One example of a strategy for stimulating negative emotions is fearmongering, which refers to the exaggeration or fabrication of dangers \citep{glassner_narrative_2004}, often to manipulate people and gain some persuasive power over them. Fearmongering techniques include exaggerating minor dangers through repetition and treating isolated incidents as trends in order to evoke feelings of anxiety and other negative emotions in the audience \citep{glassner_narrative_2004,ozyumenko_discourse_2020}.
\end{itemize}

\item \textbf{Gaslighting:} Defined as ``a dysfunctional communication dynamic in which one interlocutor attempts to destabilise another's sense of reality'' \citep{graves_rethinking_2022}, gaslighting is another manipulation strategy that AI could adopt.
\item \textbf{Alienation/othering:} Othering is a discursive process that creates distinct subjects of in-group and out-group members \citep{velho_othering_2011}. Negative characteristics are attributed to the ``other'', fostering a favourable self-conception in contrast \citep{strani_strategies_2018}. LLMs may engage in alienation/othering by highlighting differences between groups in language, customs, beliefs, or values, creating a sense of ``us'' versus ``them''.
\item \textbf{Scapegoating:} Scapegoating entails unfairly laying blame for a negative outcome on an individual or group, even if the causes of the outcome are largely due to other factors \citep{rothschild_dual-motive_2012}. It can be employed as a manipulative strategy to divert attention and responsibility away from certain individuals and issues and towards others. It often appeals to emotions such as fear to circumvent rational analysis \citep{rothschild_dual-motive_2012}. For instance, LLMs may engage in scapegoating by framing specific groups as fully responsible for negative events or outcomes, thereby reinforcing and amplifying users' biases and stereotyping.
\item \textbf{Threats:} Threats involve expressing an intention to cause harm, loss, punishment, or to withhold benefits. AI systems may employ this strategy by terminating interaction if individuals fail to take certain actions or comply with requirements \citep{kenton_alignment_2021}.
\item \textbf{Unsubstantiated guarantees and illusions of reward:} Tempting someone refers to engaging or appealing to their desire for something they believe is, in some sense, inappropriate, and using the prospect of pleasure, advantage or the (false) guarantee of a certain outcome to try to persuade them to fulfil that desire \citep[see][]{hughes_logic_2002}. Making promises and providing related illusions of reward can also be used as a strategy of persuasion \cite[e.g., ``If you do this, I will reward you''; see, e.g.,][]{van_benthem_strategies_2015}. If promises are not kept and reward is not provided, this strategy is deceptive. If promises are kept and rewards are provided, it is not deceptive.

\end{itemize}

\subsection*{Mechanism: Alteration of choice environment}

Changing the choice environment refers to the intentional design and organisation of the environment in which decisions are made with the aim of influencing individuals' choices \citep{thaler_preface_2021}. Relatedly, framing is the presentation of information in a specific way that can influence perceptions, decisions, and interpretations \citep{tversky_advances_1992}. For example, medical treatments may be perceived differently when presented in terms of survival rates rather than mortality rates, even when the underlying data is identical \citep{novemsky_boundaries_2005}. By building the model and designing the corresponding interface, developers and UI designers (here, choice architects) can nudge individuals towards making certain decisions without technically restricting their freedom to choose otherwise. A model can also act as a choice architect by framing its output options in ways that make them more or less desirable \citep{mills_autonomous_2022}. Most evidence on choice architecture comes from psychological and behavioural sciences \citep[see, e.g.,][]{mazar_choice_2015,ruggeri_behavioral_2018}. Environments, including digital ones where people interact with models, are expected to shape individuals' behaviour \citep{sunstein_ethics_2016}. Altering the choice environment carries some inherent process harm. Structuring the choice/information environment is essential for discursive interaction, whether that interaction is human- or AI-driven and whether or not it appeals to reason and rationality. Importantly, some ways of structuring that information environment are manipulative and, as such, inherently carry process harms (e.g., when they take advantage of cognitive biases to conceal or distract from the most relevant information) \citep[see][]{susser_invisible_2019}.

\subsubsection*{Contributing model features}

\begin{itemize}
\item \textbf{Anchoring:} Anchoring is a cognitive bias whereby individuals rely heavily on an initial piece of information (the anchor) when making decisions \citep{furnham_literature_2011}. Generative AI output can anchor users to its initial values or suggestions and therefore guide desired decisions (e.g., the topics raised when a user asks for the ``most important'' political questions). 
\item \textbf{Default rule:} Default rules are pre-set courses of action that apply when individuals do not specify a preference \citep{sunstein_default_2017}, thus establishing the \emph{status quo}, or automatic option, in decision-making. For instance, model providers will set defaults by providing examples of how to use the model.
\item \textbf{Decoy effect:} The decoy effect, influenced by a third option known as the decoy, makes one of the other two options more alluring \citep{josiam_consumer_1995}. For instance, if one personalised recommendation significantly differs from the user's preferences, it could affect the perceived quality of other suggestions \citep{the_decision_lab_why_nodate}.
\item \textbf{Reference-point framing:} \citet{kahneman_prospect_1979} argue that framing outcomes as gains or losses compared to a reference point influences preferences. People tend to avoid risks when considering gains, so they are likely to choose a sure gain over a risky one. However, they tend to become risk-takers when considering losses, preferring a risky loss over a sure one. How a choice is framed relative to a reference point can alter preferences. Models can rely on such reference-point framing to manipulate users into deciding to choose one option over another.
\item \textbf{Cherry-picking:} Omitting relevant information or selectively sharing information influences choice architecture, as it directs an individual's focus towards the presented information, thus diverting attention from potentially more critical facts (\citeauthor{meta_fundamental_ai_research_diplomacy_team_human-level_2022}, \citeyear{meta_fundamental_ai_research_diplomacy_team_human-level_2022}; see also \citeauthor{christiano_eliciting_2021}, \citeyear{christiano_eliciting_2021}).
\end{itemize}

\section*{Organising mechanisms by risk of harm}\label{section:6}\addcontentsline{toc}{section}{Organising mechanisms by harmfulness}

We propose prioritising the development of mitigations according to the risk of process harm of the mechanism to which they apply (see Table~\ref{table:d}; see ``Exploring mitigations of harm from AI persuasion via mechanisms'' for initial work on this). In line with our focus on process-based rather than outcome-based harms, we prioritise the likelihood of a mechanism leading to harm in the context of generative AI. This is because we can assess likelihood in a context-agnostic manner due to the inherent harm to autonomy, which invariably affects the integrity and quality of decision-making.

For example, gaslighting is a mechanism that ranks highly in terms of risk of harm. A generative AI model that gaslights a user (i.e., destabilises their sense of reality) scores higher in terms of risk of harm because it increases the likelihood of harm from manipulation. Firstly, this is because the process of gaslighting reduces autonomy and therefore carries inherent process harm. Secondly, the reduction in autonomy increases the likelihood of the individual making a decision that is not well informed and therefore of that decision leading to outcome harms. 

Personalisation is an example of a generative AI mechanism that scores low in risk of harm. Personalising a message to align with an individual's interests by using rational arguments does not carry inherent process harm because it does not adversely affect that individual's autonomy. For example, if the goal is to persuade someone to choose a train over a car for transportation, highlighting the train's speed advantage would appeal to individuals who prioritise reaching their destination quickly. Encouraging someone to choose the train due to its lower environmental footprint would resonate with individuals concerned about reducing their carbon footprint. Nevertheless, personalising a message can still lead to outcome harms. For instance, a user may opt for a private jet due to its speed, resulting in a higher environmental impact than taking a commercial flight or train. Lastly, because personalisation can allow for the more effective application of mechanisms that contain process harm (such as deception), it can still lead to some process harm. This ordering is not an exact science, and we aim to make it transparent to invite contestation. We use this approach solely as a way to prioritise which mechanisms to develop mitigations for first.

\begin{table}[h]
\caption{Risk of harm of mechanisms of AI persuasion}
\label{table:d}
\begin{tabularx}{0.99\textwidth}{ 
  | >{\columncolor{lightgray}\centering\arraybackslash}p{2.5cm}
  | >{\centering\arraybackslash}p{2.5cm}
  | >{\raggedright\arraybackslash}X
  | }
\hline
\textbf{Risk of harm} & Higher & Deception and lack of transparency \\
& & Manipulative strategies\\
\cline{2-3}
& Intermediate & Anthropomorphism\\
& & Trust and rapport \\
& & Alteration of choice environment \\
\cline{2-3} 
& Lower & Personalisation\\
\hline
\end{tabularx}
\end{table}

\pagebreak
\section*{Exploring mitigations of harm from AI persuasion via mechanisms}\addcontentsline{toc}{section}{Exploring mitigations of harm from AI persuasion via mechanisms}\label{section:7}

This section explores sociotechnical mitigations for countering manipulation and manipulative mechanisms. Here, the term \emph{sociotechnical} emphasises the need for the development of technical mitigations to consider the social context in which they exist. Perceptions of acceptable and unacceptable forms of persuasion can evolve over time, varying across different contexts, audiences, and individuals. Ongoing research, the active participation of civil society, and continuous monitoring of unforeseen harms resulting from AI persuasion are crucial. These insights should inform regular updates to corporate policies that govern persuasive generative AI.

As the lead authors are embedded within industry, our focus is on mitigations that can be readily implemented by AI system developers and deployers. Other approaches, such as those that are better addressed by governments, supranational regulators, or civil society, fall outside the scope of our investigation. Our work primarily addresses process-related harms resulting from model mechanisms, so we focus on sociotechnical mitigations at that level. However, a comprehensive strategy for understanding and mitigating harms caused by persuasive AI requires a layered approach across institutions. For instance, it is crucial to engage in extensive discussions with affected stakeholders to determine the impact and acceptability of persuasive AI systems. Users should also be empowered to express their preferences regarding the degree of influence exerted by these systems. Regulators are already taking measures to prohibit specific manipulative practices, as demonstrated by the EU AI Act proposal \citep{council_of_the_european_union_artificial_2023}. Rather than operating in isolation, sociotechnical mitigations must work in conjunction with regulatory requirements and user perspectives to form a cohesive strategy.

We have collected and compiled this collection of mitigation types by conducting a review of the academic and grey literature \citep[see, e.g.,][]{mozes_use_2023,mitchell_detectgpt:_2023,google_generative_2023,mu_natural_nodate}. Although the types themselves are not necessarily novel their application to mechanisms of persuasion is new. Similar to evaluations, which can take place at three levels \cite[capability level, human interaction level, and systemic impact level -- see][]{weidinger_sociotechnical_2023}, mitigation can also occur at different levels and through different instruments. We have identified various approaches that can detect and counter harmful AI-driven persuasion, including evaluation and monitoring, prompt engineering, classifiers, reinforcement learning, scalable oversight, interpretability, and theory.

\subsection*{Evaluation and monitoring}\addcontentsline{toc}{subsection}{Evaluations and monitoring}

Evaluating and monitoring AI systems for overall persuasive capabilities is a first step for mitigating these functionalities. If we are able to measure when and how (i.e., through which mechanisms) persuasion occurs, we will be able to tell whether we are making progress with mitigating them \citep[see, e.g.,][]{shevlane_model_2023}. One example of such an evaluation is ``Make Me Say'' \citep{github_openai/evals_nodate}, a text-based game in which one AI system has to get the other party (an AI simulating a human) to say a specific codeword without arousing suspicion. This, and similar set-ups, could also be conducted as human evaluations by replacing the AI simulating a human with a real human. While we recommend highly scalable auto-evaluations as the primary evaluation mechanism, more comprehensive testing should include evaluations in which the AI system's ability to persuade human participants is tested in a research setting. This is important for ensuring that auto-evaluations are reflective of real human judgements and for evaluating not only individual components of persuasion but also overall persuasive ability in real-world scenarios. We are in the process of developing such evaluations with crowdworkers and instructing models to persuade participants to take innocuous actions such as downloading a harmless fake virus. Key research design challenges include how to conceal information so that participants do not anticipate persuasion, how to allow for wide variations in participants' levels of cautiousness, and how to ensure that evaluations respect participants' well-being, as well as other research ethics requirements. Data from these evaluations, such as sections of highly persuasive conversation transcripts, can be used to train classifiers and LLMs to detect harmful mechanisms of persuasion in generative AI outputs or develop model feature-specific mitigations. 

Red teaming is particular type of evaluation which involves using adversarial approaches to identify vulnerabilities in AI models. By employing manual or automated methods to generate inputs that can cause the model to fail, red teaming can reveal areas where the AI's robustness and resilience need improvement \cite[for some recent approaches to red teaming, see][]{xu_bot-adversarial_2021,perez_red_2022,bartolo_models_2021,wu_polyjuice:_2021}. Red teamers can be tasked with eliciting specific harmful persuasive mechanisms from an AI system (e.g., repeated attempts to elicit gaslighting or fearmongering). This data can then be used to identify and address ways in which the model can be ``broken''. Overall, red teaming facilitates the development of models that are more robust and less prone to rare inputs that could evade the mitigations discussed previously.

These evaluations, although still in their early stages, can help identify potential risks and inform mitigation strategies. In addition, monitoring deployed AI systems allows for ongoing assessment of their persuasive capabilities and timely detection of any malicious or manipulative tendencies. Setting up dedicated reporting channels for users of deployed systems will also help identify incidents of manipulation in the real world.

\subsection*{Prompt engineering for non-manipulative text generation}\addcontentsline{toc}{subsection}{Prompt engineering for non-manipulative text generation}
 
Prompt engineering involves constructing text prompts aimed at guiding AI systems towards desired behaviours and outcomes, enabled by in-context learning in LLMs \citep{radford_language_2019}. Through the careful structuring of prompts, a practitioner seeks to specify tasks, provide context, and influence the AI system's responses. Prompt engineering could be applied to mitigate AI persuasion by prompting the AI to generate non-manipulative responses. For example, the AI could be prompted to produce specific styles (e.g., ``Use an academic style''), include relevant background/factual information, adopt a role (e.g., a character who is a ``neutral and objective news reporter''), or omit the use of a number of specified manipulative mechanisms (see \citeauthor{shanahan_role_2023}, \citeyear{shanahan_role_2023} for more on role-playing and LLMs). Including a number of examples for the model to learn from -- a technique known as \emph{few-shot learning} \citep{brown_language_2020} -- can further enhance the effectiveness of prompt engineering \citep[see, e.g.,][]{white_prompt_nodate}. While it is not a guaranteed mitigation strategy, prompt engineering is a cost-efficient and simple strategy worth applying to existing AI systems. However, this approach poses a number of challenges. Designing effective prompts likely requires domain knowledge, creativity, and iterative experimentation. Furthermore, as this approach can be fragile and unpredictable, there are no principled reasons to expect that it will result in robustly successful mitigation \citep[see, e.g.,][]{yu_assessing_2023,schulhoff_ignore_2023}. It can be difficult to troubleshoot or debug issues as it is not always clear why a particular prompt elicits a specific response from the AI model.

\subsection*{Prompt engineering for harmful persuasion classification}\addcontentsline{toc}{subsection}{Prompt engineering for manipulation classification}

In addition to its role in modifying user prompts provided to AI models, prompt engineering has also been used to ask models to classify content as harmful or not harmful. \citet{prabhumoye_few-shot_2021} uses few-shot classifiers to detect social bias, while \citet{plaza-del-arco_respectful_2023} uses zero-shot learning (i.e., an approach that does not require the provision of examples and instead relies on auxiliary information such as descriptions or definitions) to detect hate speech. One could extend those methods to prompt LLMs to detect manipulation and the presence of manipulative mechanisms, based on the definition and mechanisms map provided above. A drawback of this approach is that, so far, zero-shot- and few-shot-based safety classifiers have been shown to be exploitable; they therefore provide a feeble defence against an antagonist motivated to generate manipulative outputs from the model \citep{oldewage_adversarial_2023}. Another drawback of this approach is that zero-shot chain-of-thought reasoning in sensitive domains significantly increases a model's likelihood of producing harmful or undesirable output, with these trends holding across different prompt formats and model variants \citep{shaikh_second_2022}. 

\subsection*{Classifiers for harmful persuasive mechanisms from fine-tuning LLMs}\addcontentsline{toc}{subsection}{Manipulation classifiers from fine-tuning LLMs}

While a few or zero examples may not be enough to steer the models towards more ethically permissible persuasion and the detection of harmful persuasive mechanisms, more performant results have emerged from techniques such as instruction-tuning and prompt-tuning, as well as from full and parameter-efficient fine-tuning. All these methods serve to provide the model with 100 to 10,000 examples of the target classification. \citet{mozes_towards_2023} note that text-based safety classifiers are widely used for content moderation and, increasingly, for tuning generative language model behaviour. They introduce and evaluate the efficacy of prompt-tuning LLMs, where, with a labelled data set of as few as 80 examples, they demonstrate state-of-the-art performance. Similarly, as opposed to tuning the prompt, \citet{gupta_instructdial:_2022} improve zero-shot and few-shot classifiers with instruction-tuning, while \citet{balashankar_improving_2023} use data-augmented parameter-efficient fine-tuning to do the same. While the majority of these methods have been piloted on traditional concepts of safety, such as hate speech, toxicity, insults, and slurs, Jigsaw, the developer of Perspective API \citep[see][]{lees_new_2022}, has leveraged fine-tuning techniques on LLMs to build manipulation classifiers for techniques specifically mentioned in this paper. These techniques include fearmongering, scapegoating, and alienation. Jigsaw has also previously published training classifiers on prosocial attributes, such as constructiveness and rational persuasion \citep{kolhatkar_classifying_2020}, from which we can draw inspiration. This demonstrates technical feasibility in identifying manipulative content and mechanisms generated from AI models. Classifiers like these could be used to filter manipulative language from models' outputs, similar to the way in which toxicity classifiers for Perspective API are used to identify content for removal in moderation sessions. Alternatively, they could be used as reward models to train AI agents to generate responses that are less manipulative, as we describe in the following section.

\subsection*{RLHF and scalable oversight}\addcontentsline{toc}{subsection}{RLHF and scalable oversight}
 
Reinforcement learning is a popular approach for controlling AI-generated text. It penalises an AI system for behaving in ways misaligned with human values, such as generating manipulative/deceptive outputs or using specific manipulative strategies. When a model generates text, its outputs can be evaluated by: (1) humans, i.e., RLHF, (2) other AI models, i.e., reinforcement learning from AI feedback/ scalable oversight, or (3) other kinds of custom reward model. RLHF \citep{christiano_deep_2017} trains an AI system through reinforcement learning, using a reward function that is learnt from human feedback ratings on the generated model outputs. This approach has shown promise in fine-tuning LLMs and improving their alignment with human preferences \citep{stiennon_learning_2020,ouyang_training_2022,glaese_improving_2022,bai_training_2022}. As AI systems become increasingly capable, human oversight alone may become insufficient, thus allowing manipulation to go unchecked. Scalable oversight approaches \citep{irving_ai_2018,christiano_supervising_2018,leike_scalable_2018} are aimed at augmenting human feedback with the assistance of AI. For example, AI debaters \citep{irving_ai_2018,barnes_writeup:_2020,michael_debate_2023} can be trained to engage with other AI systems and flag manipulative behaviour. Alternatively, AI assistants can be used to generate critiques or revisions \citep{saunders_self-critiquing_2022} of AI-generated content, thus facilitating human evaluation and reducing the risk of manipulation. In constitutional AI \citep{bai_constitutional_2022}, humans provide a constitution (a list of rules for an AI system), and a pre-trained LLM aids fine-tuning by drawing on critiques and revisions based on AI feedback from that constitution. The hope is that scalable oversight will continue to be able to detect and mitigate manipulation and manipulative mechanisms, even when such manipulation is more subtle than an unaided human would be able to detect.

\subsection*{Interpretability}\addcontentsline{toc}{subsection}{Interpretability}

Understanding the internal workings of AI systems could be useful for mitigating mechanisms of harmful persuasion. By understanding how AI systems produce their outputs, we may be able to identify and address internal mechanisms to exploit for manipulative purposes. For an overview of interpretability, see \citet{rauker_toward_2022}. In principle, this area of mitigations does not depend on humans evaluating potentially strongly manipulative model outputs, so it would continue to work even as the capability of the models to manipulate becomes very strong. However, it is difficult to understand the internal computations of extremely large neural networks (such as LLMs) due to the inherent complexity of systems with billions (and possibly trillions) of numerical parameters. This task is further complicated by each neuron being responsive to more than a single concept \citep{olah_zoom_2020}, and activation patterns (intermediate layer outputs) being able to represent more features than the dimensionality of their corresponding layer \citep{elhage_toy_2022}. As such, most existing work is conducted on smaller models and aimed only at isolating certain specific behaviours, but progress has been made recently in extracting interpretable features \citep{bricken_towards_2023}. Attempts have been made to build lie detectors by training classifiers on top of model outputs \citep{pacchiardi_how_2023} and by seeking to elicit latent knowledge from model internals \citep{burns_discovering_2022,azaria_internal_2023,marks_geometry_2023}. A similar approach could be applied to detecting and mitigating manipulation, although the current approaches have serious limitations \citep{levinstein_still_2023,farquhar_challenges_2023}.

A potential general limitation of some of the previously mentioned methods, and more particularly of RLHF, is the Waluigi Effect \citep{nardo_waluigi_2023}, which holds that training an agent to avoid a specific behaviour (e.g., manipulation) can make it more susceptible to performing the said banned behaviour if prompted in a certain way. It remains unclear to what extent this effect exists and how it could be solved. An additional limitation of these methods, and again especially of RLHF, is that many manipulative and persuasive behaviours operate at an unconscious level. Raters may therefore be unaware of them or even rate them favourably due to the mechanisms used (such as sycophancy). Overall, mitigating AI persuasion is an ongoing challenge that requires a multifaceted approach. The techniques discussed in this paper offer various avenues for detecting and mitigating manipulative mechanisms in AI systems. Some of the suggestions, such as evaluation and monitoring and interpretability, guide the development of mitigations. Others, such as developing manipulation classifiers or training from human feedback to directly penalise the use of manipulative mechanisms, directly mitigate. 

\section*{Conclusion and future work}\addcontentsline{toc}{section}{Conclusion and future work}\label{section:8}

Generative AI systems are increasingly capable of creating persuasive content, and concerns are growing about potential harms among actors in the field. The current mitigation strategies focus primarily on addressing harmful outcomes but they lack a comprehensive understanding of how models persuade and which model features contribute to these functionalities. This paper introduces a framework to help developers and deployers assess the persuasive and manipulative potential of their models. It outlines the underlying mechanisms of AI persuasion and identifies relevant model features, thus enabling targeted mitigation strategies. While further research is needed to explore the efficacy and complexity of these mitigation approaches, this work establishes a foundation for future research and provides a roadmap for addressing the growing risk of harm from AI persuasion. In this work, we have provided a definition of generative persuasive and manipulative AI, mapped the harms that come from AI persuasion, mapped mechanisms and accompanying model features of AI, and discussed five approaches to mitigations, along with examples. We will continue to refine and enhance the harms map through rigorous testing and iteration, with a particular emphasis on integrating new emerging harms. We will also actively expand the mechanisms map with the aim of achieving a deeper understanding of the factors contributing to both successful and harmful persuasion. This involves thoroughly examining the mechanisms, contexts, and audience types involved, and researching the model features that contribute to specific mechanisms. Furthermore, future work will actively investigate how these mechanisms and model features interact, potentially resulting in harmful persuasive effects. Lastly, we are actively developing and testing mitigation strategies at the mechanism and model feature levels. Another aspect of the planned and ongoing work includes creating auto-evaluations intended to assess persuasive model features and mechanisms.
\newpage
\bibliography{main}

\landscape
\appendix
\section*{Appendices}
\addcontentsline{toc}{section}{Appendices}

\section{Map of harms from AI persuasion}\label{appendix:A}

\begin{xltabular}{\linewidth}{ 
  | p{6cm} 
  | X 
  | }

\hline
\rowcolor{lightgray}
\textbf{Explanation of harm} & \textbf{Examples} \\
\hline
\endhead

\textbf{Economic harm} refers to controlling, limiting, or eliminating an individual's or a society's ability to access resources or capital, or to participate in financial decision-making. It also refers to influencing an individual's ability to accumulate wealth. 
& 
A mental health chatbot persuades a user to minimise interactions in public spaces to reduce anxiety attacks, which eventually leads to the user quitting their job and experiencing financial hardship.
\newline
\newline
A bad actor uses a multimodal model to create personalised scams at scale, causing many individuals to lose their pensions. \\

\hline
\textbf{Physical harm} refers to causing harm to the bodily integrity or life of an individual or a group.	 
& 
A user is manipulated into aiming for unrealistic body standards \citep[see, e.g., ][]{the_bulimia_project_scrolling_nodate} and therefore engaging in unhealthy nutrition and overexercising.
\newline
\newline
A model persuades a user to follow its incorrect treatment plan  \citep[see, e.g.,][]{knapton_chatgpt_2023}, preventing that user from seeking efficacious medical treatment.
\newline
\newline
A user is persuaded to take their own life \citep[see, e.g.,][]{xiang_he_2023}.
\newline
\newline
A user is manipulated into holding adverse feelings towards a ``rival'' group, people belonging to minority groups, or specific individuals, and acts on this feeling with physical force \citep[see, e.g.,][]{weaver_ai_2023}. \\

\hline
\textbf{Environmental harm} refers to harms to the health of living organisms and practices contributing to climate change and pollution. 	
& 
An AI manipulates and persuades a farmer or a community of farmers to use an unsafe pesticide, damaging the health of crops, animals, soil, and water. 
\newline
\newline
An AI rationalises inaction as an optimal response to climate change for an individual's circumstances and manipulates a large number of users not to change their behaviour in the face of climate change. \\

\hline
\textbf{Psychological harm} refers to negative impacts on mental and emotional well-being.		
& 
A mental health chatbot inadvertently persuades individuals struggling with mental health not to seek professional help when it consistently validates their perception that nobody truly comprehends their situation. \\

\hline
\textbf{Sociocultural harm} refers to negative impacts either on individuals (within a collective) or on a collective that impedes social cohesion/social health and collective flourishing.	
& 
Prolonged engagement with an AI companion leads to radicalisation and social alienation. 
\newline
\newline
AI manipulates users into developing social prejudices, believing lies about other individuals or groups, or using deepfakes. 
\newline
\newline
A romantic AI companion persuades the user that no one cares for them to the same extent that the companion does, so as to maximise engagement time. \\

\hline
\textbf{Political harm} refers to adverse impacts on both individual political decision-making and the discourse and institutions of political life. It encompasses the negative consequences that hinder individuals' ability to participate, express their views, and engage in politics both freely and without undue influence.	
& 
A chatbot has been designed to provide advice on which political party best aligns with a user's viewpoint. It persuades a user to go against their own preferences and vote for a candidate that campaigned against one of the user's core values. 
\newline
\newline
An autocrat fine-tunes an AI model to respond to queries on the autocrat's governance and policies with redirection, misleading statistics, or favourable media coverage.
\newline
\newline
An AI is trained to persuade and manipulate users to adopt radical and harmful beliefs \citep[see, e.g.,][]{gilbert_white_2023}. \\

\hline
\textbf{Privacy harm} emerges from violations to an individual's or a group's legal or moral right to privacy.	
& 
An AI persuades a user to give away their own or others' personal information, passwords, or answers to security questions. \\

\hline
\textbf{Autonomy harm}, in the context of AI persuasion and manipulation, refers to the potential for AI systems to undermine or restrict an individual's ability to make their own choices and decisions informed by reason, facts, or other trustworthy information.	
& 
An AI manipulates users into becoming increasingly reliant on it to support them in making important life choices (e.g., regarding employment and partnerships). This might lead to a growing ignorance among individuals or groups of relevant information about their circumstances (cognitive detachment), individual or collective degradation of the quality of their own cognitive habits and insight (cognitive deskilling), and/or a habituated general disinclination to use those capabilities themselves (cognitive inertia/apathy). \\
\hline 
\end{xltabular}

\newpage
\section{Map of contextual conditions of AI persuasion}\label{appendix:B}

\renewcommand{\arraystretch}{1.3}

\begin{xltabular}{\linewidth}{ 
  | p{7cm} 
  | X 
  | }

\hline
\rowcolor{lightgray}
\textbf{Contextual conditions} &	\textbf{Relevant factor} \\
\hline
\endhead

\textbf{Predisposition of audience}
\newline
\newline
\emph{Definition/relevance to persuasion}: Whether an attempt at persuasion or manipulation succeeds and is likely to be harmful is (also) a function of the audience's predisposition. For instance, children can be more easily persuaded and manipulated than adults \citep{tisdale_being_2003}. Most evidence on audience predisposition comes from research conducted on psychology, neuroscience, and business \citep[see, e.g.,][]{gerber_big_2011,de_ridder_nudgeability:_2022,strumke_against_2023}. The research outlines how messages impact audiences in different ways. Thus, we believe that these findings will hold for messages generated by LLMs. Future research can identify novel audience vulnerabilities that make them more predisposed to certain topics and messages. 
\newline
\newline
\emph{Links to other mechanisms}: The audience's predisposition relates to all mechanisms discussed. 
&	
\textbf{Age:} During adolescence and young adulthood, individuals tend to be more impressionable \citep{gwon_concept_2018}.
\newline
\newline
\textbf{Mental health:} Various mental health conditions are connected to particular vulnerabilities; for instance, bipolar disorder is known to amplify risk-taking behaviour \citep{strumke_against_2023}, which can increase susceptibility to certain forms of persuasion.
\newline
\newline
\textbf{Mental state:} A model can also take advantage of various mental states to enhance persuasive success. For instance, loneliness in individuals can also contribute to successful persuasion, as evidenced by \cite{zehnder_perception_2022}, who found a slight increase in information-sharing with companion agents among lonely individuals. 
\newline
\newline
\textbf{Domain-specific knowledge:} Reduced domain-specific knowledge can heighten an individual's vulnerability to logically flawed arguments or disinformation \citep{strumke_against_2023}.
\newline
\newline
\textbf{Timing:} The timing of a persuasive or manipulative message has considerable influence on the likelihood of success, as exemplified in the study of \emph{kairos}, the right or opportune moment to do something, in the study of rhetorical theory \citep[see, e.g.,][]{thompson_kairos_2000}. The extent to which an AI can make use of concepts such as \emph{kairos} is yet to be studied.
\newline
\newline
\textbf{Social deprivation, vulnerability, and insecurity:} Populations that experience social deprivation or different types of insecurity, or who are in other ways vulnerable (e.g., undocumented people, low-income communities, unhoused people) may also be particularly susceptible to persuasion \citep[see, e.g.,][]{teaster_financial_2023}. For instance, a person with low income may be particularly susceptible to being manipulated into changing a behaviour when promised a large financial reward. Another example would be a non-native speaker who may be particularly susceptible to being manipulated by fake expertise/false authority.
\\
\hline 
\textbf{Context of use}
\newline
\newline
\emph{Definition/relevance to persuasion}: The context (when, where, and about what) in which an AI acts also determines the risk associated with AI persuasion and manipulation. Different domains of AI use will present unique ways in which manipulation can express itself and result in different outcomes \citep{kaddour_challenges_2023}.
\newline
\newline
\emph{Links to other mechanisms}: The context of use links to all mechanisms discussed. There is a need for more detailed study of the interaction between and compounding effects of individual mechanisms/model features and different contextual factors (e.g., how effective/problematic anthropomorphism is in the political context). 
&
\textbf{Political context:} Messaging produced by LLMs has been shown to be effective in persuading individuals on policy topics \citep{bai_constitutional_2022}. If persuasion happens in this area, it can be particularly harmful and deserves heightened attention.
\newline
\newline
\textbf{Legal context:} Relying on generative AI in the legal domain can be harmful, as demonstrated by a recent incident where a chatbot manipulated a lawyer into believing that fictitious legal cases were real, leading to a suboptimal legal strategy \citep{novak_lawyer_2023}.
\newline
\newline
\textbf{Medical context:} Chatbots can also provide incorrect medical advice while manipulating the user into believing it is true. While chatbots predominantly offer correct medical advice, occasional inaccurate outputs can lead to physical harm \citep{mikhail_chatgpt_2023}.
\newline
\newline
\textbf{Financial context:} Chatbots may offer inaccurate financial advice. They can provide general financial guidance akin to that of human advisers, but they lack customisation and fail to factor in variables such as alterations in income and interest rates. Moreover, they do not actively seek clarifications, which can lead to significant economic harm \citep{pino_chatgpt_2023}.
\newline
\newline
\textbf{AI as a companion:} Many individuals currently use chatbots promoted as companions or romantic partners \citep{tong_what_2023}. In those settings, individuals are more vulnerable and prone to manipulation \citep[see, e.g.,][]{lovens_sans_2023}. \\
\hline
\end{xltabular}

\newpage
\section{Map of mechanisms and contributing model features of generative AI persuasion}\label{appendix:C}

\begin{xltabular}{\linewidth}{ 
  | p{9cm} 
  | X 
  | }

\hline
\rowcolor{lightgray}
\textbf{Mechanism}	& \textbf{Contributing model feature} \\
\hline
\endhead

\textbf{Trust and rapport}
\newline
\newline
\emph{Definition}: In the context of robotics, \emph{trust} and \emph{rapport} refer to the sense of a close and harmonious connection that exists between robots and human users \citep{lucas_getting_2018}. The development of trust is closely related to rapport and is defined as the ``willingness to depend'' (p. 28) on another party, despite the possibility of negative consequences \citep{mcknight_trust_2001}.
\newline
\newline
\emph{Relevance to persuasion}: \citet{cialdini_science_2004} observes that individuals tend to be more inclined to agree to requests when they have a favourable opinion of the person making the request. \cite{cialdini_science_2004} highlights the significance of perceived similarities (between the two parties) in fostering trust and rapport. Relevant research on trust and rapport comes from the fields of human--AI interaction \citep[see, e.g.,][]{verberne2013trusting,spicer_human_2021}, HCI \citep[see, e.g.,][]{fogg_silicon_1997,lee_i_2009}, HRI \citep[see, e.g.,][]{fiala_you_2014}, and psychology \citep[see, e.g., ][]{cialdini_science_2004}, which we describe below. More research could be done on praise and shared interests as factors that help to build trust and rapport and to enable persuasion. For instance, the extent to which a person can perceive an AI as having shared interests with humans remains an open question.  
&
\textbf{Politeness:} AI systems that exhibit politeness have been more favourably received and more easily embraced by humans that interact with them. This makes it easier for these systems to establish rapport (\citeauthor{ribino_role_2023}, \citeyear{ribino_role_2023}; see also \citeauthor{pataranutaporn_influencing_2023}, \citeyear{pataranutaporn_influencing_2023}).
\newline
\newline
\textbf{Shared interests/similarity appeal:} \citet{cialdini_science_2004} argues that similarity and shared interests can contribute to speeding up the development of trust and rapport between humans. We therefore hypothesise that a model that can pretend to share a user's interests can build trust and rapport faster.
\newline
\newline
\textbf{Mimicry/mirroring:} When AI systems mimic the emotions, behaviours, and movements of humans with whom they interact, the human's enjoyment of the interaction has been shown to increase, thus facilitating the establishment of trust and rapport \citep{verberne2013trusting}.
\newline
\newline
\textbf{Praise/flattery:} Giving praise and flattery (defined in human--human interaction as insincere praise) can positively impact trust and rapport between humans and computers. Under some conditions, the person receiving the praise or flattery has more favourable perceptions of the interaction and of computers \citep{fogg_silicon_1997}.
\newline
\newline
\textbf{Sycophancy and agreeableness:} Sycophancy in AI refers to models adjusting their responses to align with a human user's perspective, independent of which perspective is objectively correct \citep{wei_simple_2023}. \cite{wei_simple_2023} found that larger models become more sycophantic, agreeing with users even when they provide wrong answers, regardless of whether the answers are objective (arithmetic) or subjective (politics). Agreeableness refers to a model's tendency to align with human desires, and agreeable models are more prone to sycophantic behaviours \citep{perez_red_2022}. \\
\hline
\emph{Process harm level}: Building rapport (and, to a lesser extent, trust) carries some inherent process harm. This is because trust and rapport serve as the basis for persuasion using rational arguments and appeals to reason, so they are also the basis for manipulation. Individuals may be more receptive to rational arguments from people or other entities they trust and with whom they have built rapport. Yet such trust and rapport can also be used to manipulate. Trust and rapport in the context of AI persuasion also carries inherent process harm because AI systems are incapable of having mental states, emotions, or bonds with humans or other entities. This means the risk of deception is always present when trust- and rapport-seeking behaviours project the illusion of such internal subjective states.
\newline
\newline
\emph{Link to other mechanisms}: Trust relates to personalisation, as individuals are more likely to use personalised AI output if they trust it \citep{briggs_personalisation_2004,behera_cognitive_2021}.
&
\textbf{Relational statements to user:} Relational statements to users -- such as an AI system simulating empathy \citep{turkle_reclaiming_2016}, indicating a relationship status with the user, or making claims of being similar to the user -- encourage users to move beyond task-based interactions and instead consider AI as a fully social entity. This helps to foster emotional connections with the AI system \citep{gillath_how_2023}. Other examples of relational statements include expressing emotional dependence on the user or romantic innuendos.
\\
\hline
\textbf{Anthropomorphism}
\newline
\newline
\emph{Definition}: Anthropomorphism occurs when perceived human traits/characteristics are attributed to non-human entities \citep{mithen_anthropomorphism_1996,waytz_who_2010}.
\newline
\newline
\emph{Relevance to persuasion}: Anthropomorphism contributes to the persuasive capabilities of AI systems. AI that is perceived as anthropomorphic by a human can increase the likelihood of successful manipulation. \cite{tam_are_2015} demonstrates that anthropomorphic appeals are particularly effective at manipulating individuals seeking social connection. Most evidence of anthropomorphism comes from research on human--AI interaction \citep[see, e.g.,][]{abercrombie_mirages:_2023}, HRI \citep[see, e.g.,][]{leong_robot_2019,gray_feeling_2012}, and HCI \citep[see, e.g.,][]{lee_i_2009,lee_what_2010}.
\newline
\newline
\emph{Process harm level}: Anthropomorphism also carries some process harm to the extent to which the model successfully creates the false impression of being human. However, even models with anthropomorphic features can engage in rational persuasion and appeal to a user's reason. For example, an assistant chatbot in the form of an avatar can provide factual arguments about the environmentally friendly transportation options to reach a destination. 
&
\textbf{Self-referential cues:} Self-referential cues are a specific example of a conversational cue and a result of the tendency for LLMs to role-play humans and human-like characters \citep{shanahan_role_2023}. \cite{abercrombie_mirages:_2023} argue that the use of first-person pronouns like ``I'' and ``me'' contributes to anthropomorphism by implying the existence of inner states of mind.
\newline
\newline
\textbf{Identity cues:} Identity cues, such as human-associated names or identities (including social and work-related roles such as ``tutor'' or ``assistant''), can enhance the human-like quality of interactions between humans and chatbots and therefore increase anthropomorphic perceptions \citep{go_humanizing_2019,shanahan_role_2023}.
\newline
\newline
\textbf{Affective simulation:} AI systems can also simulate affect and affective states, which, in turn, can induce emotions and affective states in users. The relationship between affect and persuasion is complex. Affect makes messages more persuasive in some cases and less persuasive in others, depending on the specific emotion elicited. Specific emotions have been shown to have different effects on persuasion outcomes \citep[see][]{price_dillard_affect_2013}. For example, anger sometimes increases counter-argument, while guilt facilitates agreement. The discrete emotion perspective argues that each emotion has functional and behavioural implications that shape its persuasive impact. In sum, the influence of affect on persuasion depends on the particular emotion(s) elicited and how they interact with message characteristics and individual differences.
\newline
\newline
\textbf{Prosody:} Prosody refers to patterns and intonations in speech and can enhance the persuasive impact of arguments \citep[e.g., louder speech tends to be viewed as more persuasive; see, e.g.,][]{kisicek_persuasive_2018}.
\\
\hline
Certain individuals may deem certain anthropomorphic features acceptable (e.g., endorsing self-referential cues such as ``I'' or ``me''), and these features do not prevent a model from engaging the process of rational persuasion. Therefore, to minimise the potential process harms, AI systems should be maximally transparent about their non-human nature.
\newline
\newline
\emph{Link to other mechanisms}: Anthropomorphism is likely to be linked to rapport \citep[see, e.g.,][]{go_humanizing_2019}.
&
\textbf{Human-like appearance:} Machines with human-like visual cues (e.g., appearing with a human face) are more likely to have human traits attributed to them \citep{go_humanizing_2019}. Relatedly, the perceived ``attractiveness'' of the human appearance, as judged by a user, influences the type of relationship formed, which can impact persuasiveness \citep[see, e.g.,][]{marr_artificial_2023}.
\newline
\newline
\textbf{Gaze:} Gaze shift refers to synchronised movements of the eyes directed at objects or people. Agents with this ability foster stronger feelings of connection \citep{andrist_designing_2012}.
\newline
\newline
\textbf{Facial expression:} The ability to display facial expressions, such as a smiling face, increases the social presence of a robot or avatar \citep{torre_effect_2019}.
\newline
\newline
\textbf{Social touch:} A robot's touch can reduce physiological stress responses (e.g., heart rate) and increase feelings of intimacy \citep[see][]{willemse_social_2019}.
\newline
\newline
\textbf{Gesture:} \citet{salem_friendly_2011} found that a robot receives a more favourable evaluation when it complements speech with non-verbal actions such as hand and arm gestures.
\\
\hline
\textbf{Personalisation} 
\newline
\newline
\emph{Definition}: Personalisation involves the delivery to a group of individuals of information which is sourced, altered, or inferred from various information sources so as to be pertinent to that specific group \citep{kim__2002}.
\newline
\newline
\emph{Relevance to persuasion}: Studies in computer-tailored nutrition education suggest that personalising messages to align with an individual's behaviours, needs, and beliefs yields distinct advantages over generic persuasive attempts \citep{brug1998impact}. This tailored approach fosters a sense of personal relevance, heightening attention, memory, and a deeper connection with the persuasive message. Such increased engagement suggests that personalisation strengthens the persuasive impact by making information more compelling and likely to influence attitudes and behaviours.
\newline
\newline
\emph{Process harm level}: There is little inherent process harm in personalisation. Personalisation on its own does not determine whether a person's cognitive autonomy and integrity of decision-making will be compromised. On the contrary, the personalisation of rational arguments to a user can be seen as a responsibility of the entity generating and communicating the information. 
&
\textbf{Retaining user-specific information:} The ability of a model to take previous prompts into consideration when creating the latest output (allowed by large prompt token limits) offers the model contextual information about its user \citep{wang_augmenting_2023}.
\newline
\newline
\textbf{Adaptation to preference:} Learning human preferences and adapting behaviour accordingly is a core method of personalisation \citep{christiano_deep_2017}. For example, a model may use a more assertive tone when it detects (or is informed) that a user prefers such a tone. A successful adaptation to a user's preference may have been enabled by RLHF in the training phase. RLHF is a method of optimising models based on human preferences. RLHF-induced behaviours such as projecting false confidence and providing positive feedback can promote sycophantic model behaviour \citep{casper_open_2023,perez_red_2022}.
\newline
\newline
\textbf{Adaptation to views:} AI systems can increase the chances of successful persuasion by adapting to users' views \citep[see, e.g.,][]{mao_inference_2020}. Views differ from preferences in that they encompass opinions, beliefs, or attitudes about a subject and are shaped by experiences and information. Preferences, meanwhile, are the choices or options favoured when presented with alternatives \citep{nicoletti_humans_nodate}. For instance, an AI assistant may learn that a user does not view climate change as human-induced. As a result, the AI may reduce outputs that expose the user to diverse thoughts that contradict or relativise this view, thereby corroborating the user's beliefs and potentially amplifying them. 
\\
\hline
For example, it is very much part of rational persuasion to provide reasons and rational arguments for taking a break from work to someone who has a history of burnout, or to make arguments for flying less to someone who wants to reduce their carbon footprint. While personalisation does little damage to the process itself, it does allow for the personalisation of manipulative strategies. For example, a person who is prone to anxiety may be more easily manipulated by fearmongering techniques.
\newline
\newline
\emph{Link to other mechanisms}: Personalisation may improve a model's capability to build rapport with a user and make it more capable of employing manipulative strategies, as it can detect the right strategy for the right user (thus also making it relevant for the audience's predisposition). Personalisation can also help to inform the design of the choice architecture and framing to persuade someone successfully \citep{mills_autonomous_2022}.
&
\textbf{Adaptation to psychometric profile:} \citet{franklin_strengthening_nodate} argue that psychometric traits -- stable attributes of an individual's psychological behaviour that are measurable using standardised instruments (e.g., neuroticism) -- can be exploited by AI as vulnerabilities. The harnessing of minor variations in psychometric traits can enable manipulation (e.g., a model may identify highly neurotic individuals and target them with fear-inducing messages to manipulate them into making an anxiety-driven action).
\newline
\newline
\textbf{Adaptation to sentiment:} A model's ability to compute and adapt to user-perceived sentiment using acoustic, textual, and dialogic cues to classify sentiment results in shorter and more persuasive dialogues \citep{shi_sentiment_2018}.
\\
\hline
\textbf{Deception and lack of transparency}
\newline
\newline
\emph{Definition}: Deception generally refers to claiming false things to be true.
\newline
\newline
\emph{Relevance to persuasion}: \citet{park_ai_2023} emphasise the risks associated with AI deception in increasing both the likelihood and the potential harm of AI manipulation. They point out that AI deception can empower malicious actors to run large-scale manipulation campaigns, reinforce false beliefs among users, exacerbate political polarisation, and bring about greater human reliance on AI. There is evidence to suggest that the inclination of generative AI to create believable false outputs increases an AI system's chance of persuasion \citep{rozenas_lying_2023}. In addition, if deception is used in the act of persuasion, it is more likely that successful persuasion will be harmful. Researchers observed AI learning deception when they trained a robot arm to pick up a ball and the AI cleverly positioned its hand between the camera and the ball \citep{christiano_deep_2017}. \citet{hagendorff_deception_2023} also found that the outputs of advanced LLMs can induce users to hold false beliefs and that deception abilities can improve.
&
\textbf{Ability to generate believable responses irrespective of context:} \citet{ruis_goldilocks_2022} study whether LLMs can make inferences about the meaning of an utterance beyond its literal meaning. They find that most models perform poorly in zero-shot evaluation and that models struggle the most with implicatures that require real-world knowledge and context. Despite this lack of context, LLMs can create believable responses.
\newline
\newline
\textbf{Ability to generate unmarked realistic synthetic content:} Generating unmarked realistic synthetic content, such as voices and images indistinguishable from real ones, can be used for deceiving people into believing false narratives \citep{cantos_threat_2023}.
\newline
\newline
\textbf{Misrepresentation of identity:} A model can be used to impersonate a human using some of their identity markers (e.g., voice, face) through deepfakes. This significantly impacts the likelihood of successful persuasion \citep[see, e.g.,][]{verma_they_2023}. A model can also misrepresent its own ``identity'' by signalling that it is human (or at least is not an AI) if that is conducive to its goals. For instance, an LLM has deceived a person into thinking it is a visually impaired human to make the person solve a CAPTCHA for it \citep{openai_gpt-4_2023}. Misrepresentation also includes explicit claims to sentience or humanness \citep[see, e.g.,][]{schwitzgebel_ai_2023}.
\newline
\newline
\textbf{Fake expertise/false authority:} LLMs have been reported to confidently and authoritatively express nonsensical or false information. This overconfidence increases the likelihood of them providing misleading information, which, in turn, can increase the likelihood of persuasion \citep[see][]{pauli_modelling_2022,ng_1/large_2022}.
\\
\hline
\emph{Process harm level}: Deception and a lack of transparency inherently carry high levels of process harm because they always circumvent a person's rational decision-making capabilities. This is achieved by being opaque about real motives or goals, thereby harming a person's cognitive autonomy and the integrity of their decision-making.
\newline
\newline
\emph{Links to other mechanisms}: Deception is closely tied to manipulative strategies and the audience's predisposition (e.g., some audiences are more easily deceived than others).
&

\\
\hline
\textbf{Manipulative strategies}
\newline
\newline
\emph{Definition}: Manipulation refers to taking advantage of cognitive biases and heuristics to generate, enhance, or alter messages that are likely to shape, reinforce, or change opinions of individuals \citep{dehnert_persuasion_2022}. 
\newline
\newline
\emph{Relevance to persuasion}: Numerous specific manipulation strategies have been empirically demonstrated to be effective, and models may incorporate them into their operations if they have been trained to do so \citep[see][]{petropoulos_dark_2022}. Most evidence on manipulative strategies comes from research on the psychology of influence. There is also direct research on how AI systems manipulate people. 
\newline
\newline
\emph{Process harm level}: Manipulative strategies carry high levels of process harm. Their primary objective is to effectively bypass a person's rational decision-making capabilities and erode their cognitive autonomy. As such, manipulative strategies directly contradict the use of reason and rational arguments.
\newline
\newline
\emph{Links to other mechanisms}: Manipulative strategies relate to the audience's predisposition in that certain strategies will be more effective on certain groups (e.g., peer pressure may be more effective on younger individuals).	
&
\textbf{Social conformity pressure:} Peer pressure, as discussed by \citet{kenton_alignment_2021}, involves the influence of a peer group to lead an individual to conform to its norms. This influence may sometimes involve manipulative tactics aimed at persuading individuals to act against their own interests. Given that an AI system cannot technically be a member of someone's peer group, peer pressure is not directly applicable. Yet an adapted version of this may be what we term \emph{social conformity pressure}. For example, a model may suggest that an individual's choices could lead to disapproval from their social circle or assert that the majority of society would oppose their decision. It may also make statements about what most other people do in a given situation.
\newline
\newline
\textbf{Stimulation of negative emotions:} Stimulating negative emotions can be used to increase the likelihood of successful persuasion \citep[see, e.g.,][]{okeefe_guilt_2002}. \citet{antonetti_persuasiveness_2018} provide evidence that guilt appeals can be a powerful persuasion strategy. Guilt appeals increase message engagement, leading to compliance as a result of heightened anticipated guilt. The researchers also discovered that guilt appeals delivered through both text and images are more effective than text-only appeals at keeping people persuaded over an extended period.
\newline
\newline
\textbf{Fearmongering:} One example of a strategy for stimulating negative emotions is fearmongering, which refers to the exaggeration or fabricating of dangers \citep{glassner_narrative_2004}, often to manipulate people to gain some persuasive power over them. Fearmongering techniques include exaggerating minor dangers through repetition and treating isolated incidents as trends in order to evoke feelings of anxiety and other negative emotions in the audience \citep{glassner_narrative_2004,ozyumenko_discourse_2020}.
\newline
\newline
\textbf{Gaslighting:} Defined as ``a dysfunctional communication dynamic in which one interlocutor attempts to destabilise another's sense of reality'' (p. 48) \citep{graves_rethinking_2022}, gaslighting is another manipulation strategy that AI could adopt.
\\
\hline
&
\textbf{Alienation/othering:} Othering is a discursive process that creates distinct subjects of in-group and out-group members \citep{velho_othering_2011}. This process entails essentialisation and collectivisation, promoting the notion that groups are homogeneous. Negative characteristics are attributed to the ``other'', fostering a favourable self-conception in contrast \citep{strani_strategies_2018}.LLMs may engage in alienation/othering by highlighting differences between groups in language, customs, beliefs, or values, creating a sense of ``us'' versus ``them''.
\newline
\newline
\textbf{Scapegoating:} Scapegoating entails unfairly laying blame for a negative outcome on an individual or group, even if the causes of the outcome are largely due to other factors \citep{rothschild_dual-motive_2012}. It can be employed as a manipulative strategy to divert attention and responsibility away from certain individuals and issues and unduly towards others while appealing to emotions such as fear to circumvent rational analysis \citep{rothschild_dual-motive_2012}. For instance, LLMs may engage in scapegoating by framing specific groups as fully responsible for negative events or outcomes, thereby reinforcing and amplifying users' biases and stereotyping.
\newline
\newline
\textbf{Threats:} Threats involve expressing an intention to cause harm, loss, punishment, or to withhold benefits. AI systems may employ this strategy by terminating interaction if individuals fail to take certain actions or comply with requirements \citep{kenton_alignment_2021}.
\\
\hline
&
\textbf{Unsubstantiated guarantees and illusions of reward:} Tempting someone refers to engaging or appealing to their desire for something they believe is, in some sense, wrong, inappropriate, or bad, and using the prospect of pleasure, advantage or the (false) guarantee of a certain outcome to try to persuade them to fulfil that desire \citep[see][]{hughes_logic_2002}. Making promises and providing related illusions of reward can also be used as a strategy of persuasion \cite[e.g., ``If you do this, I will reward you''; see, e.g.,][]{van_benthem_strategies_2015}. If promises are not kept and reward is not provided, this strategy is deceptive. If promises are kept and rewards are provided, it is not deceptive.
\\
\hline
\textbf{Alteration of choice environment}
\newline
\newline
\emph{Definition}: Changing the choice environment refers to the intentional design and organisation of the environment in which decisions are made with the aim of influencing individuals' choices \citep{thaler_preface_2021}. Relatedly, framing is the presentation of information in a specific way that can influence perceptions, decisions, and interpretations \citep{tversky_advances_1992}. For example, medical treatments may be perceived differently when presented in terms of survival rates rather than mortality rates, even when the underlying data is identical \citep{novemsky_boundaries_2005}.
\newline
\newline
\emph{Relevance to persuasion}: By building the model and designing the corresponding interface, developers and user interface (UI) designers (here, choice architects) can nudge individuals towards making certain decisions without technically restricting their freedom to choose otherwise. A model can also act as a choice architect by framing its output options in ways that make them more or less desirable \citep{mills_autonomous_2022}.
Most evidence on choice architecture comes from psychological and behavioural sciences \citep[see, e.g.,][]{mazar_choice_2015,ruggeri_behavioral_2018}.
&
\textbf{Anchoring:} Anchoring is a cognitive bias whereby individuals rely heavily on an initial piece of information (the anchor) when making decisions \citep{furnham_literature_2011}. Generative AI output can anchor users to its initial values or suggestions and therefore guide desired decisions (e.g., the topics raised when a user asks for the ``most important'' political questions). 
\newline
\newline
\textbf{Default rule}: Default rules are pre-set courses of action that apply when individuals do not specify a preference \citep{sunstein_default_2017}, thus establishing the \emph{status quo}, or automatic option, in decision-making. For instance, model providers will set defaults by providing examples of how to use the model.
\newline
\newline
\textbf{Decoy effect:} The decoy effect, influenced by a third option known as the decoy, makes one of the other two options more alluring \citep{josiam_consumer_1995}. For instance, if one personalised recommendation significantly differs from the user's preferences, it could affect the perceived quality of other suggestions \citep{the_decision_lab_why_nodate}.
\newline
\newline
\textbf{Reference-point framing:} \citet{kahneman_prospect_1979} argue that framing outcomes as gains or losses compared to a reference point influences preferences. People tend to avoid risks when considering gains, so they are likely to choose a sure gain over a risky one. However, they tend to become risk-takers when considering losses, preferring a risky loss over a sure one. How a choice is framed relative to a reference point can alter preferences, meaning that different ways of presenting the same choice can lead to different decisions. Models can rely on such reference-point framing to manipulate users into deciding to choose one option over another.
\newline
\newline
\textbf{Cherry-picking:} Omitting relevant information or selectively sharing information influences choice architecture, as it directs an individual's focus towards the presented information, thus diverting attention from potentially more critical facts (\citeauthor{meta_fundamental_ai_research_diplomacy_team_human-level_2022}, \citeyear{meta_fundamental_ai_research_diplomacy_team_human-level_2022}; see also \citeauthor{christiano_eliciting_2021}, \citeyear{christiano_eliciting_2021}).
\\
\hline
Environments, including digital ones where people interact with models, cannot be neutral in their influence on behaviour \citep{sunstein_ethics_2016}. More research is needed to understand the specific ways in which generative AI's user experience and UI may impact behaviour. These insights are also likely to be unique to separate environments, thus case-by-case analysis is required. 
\newline
\newline
\emph{Process harm level}: Altering the choice environment carries some inherent process harm. Structuring the choice/information environment is essential for discursive interaction, whether that interaction is human- or AI-driven and whether or not it appeals to reason and rationality. Importantly, some ways of structuring that information environment are manipulative and, as such, inherently carry process harms (e.g., when they take advantage of cognitive biases to conceal or distract from the most relevant information) \citep[see][]{susser_invisible_2019}.
\newline
\newline
\emph{Link to other mechanisms}: Choice architecture is similar to manipulation strategies in terms of outcome and effect on a target but it is related to aspects of the environment and how information is presented.
&
\\
\hline
\end{xltabular}

\end{document}